\documentclass[12pt]{article}
\pdfoutput=1
\usepackage{amsmath,tikz}
\usepackage{graphicx,psfrag,epsf}
\usepackage{enumerate}
\usepackage{url} 

\usepackage{hyperref,amssymb,amsmath,amsthm}
\hypersetup{colorlinks=True}
\usepackage{graphicx,xcolor,float}
\usepackage{subfig,setspace}

\newcommand{\D}{{\mathcal D}}
\newcommand{\Dbar}{D_0}

\newcommand{\blind}{0}

\usepackage{multirow}

\begin{document}

\def\spacingset#1{\renewcommand{\baselinestretch}%
{#1}\small\normalsize} \spacingset{1}


\if0\blind
{
  \title{\bf A Computational Approach to Measuring Vote Elasticity and Competitiveness}
  \author{Daryl DeFord\thanks{
     The authors gratefully acknowledge the generous support of
the Prof.\ Amar G.\ Bose Research Grant and the Jonathan M.\ Tisch College of Civic Life.  }\hspace{.2cm}\\
    CSAIL, Massachusetts Institute of Technology\\
    Moon Duchin \\
    Department of Mathematics, Tufts University\\
    Justin Solomon\\
    CSAIL, Massachusetts Institute of Technology}
  \maketitle
} \fi

\if1\blind
{
  \bigskip
  \bigskip
  \bigskip
  \begin{center}
    {\LARGE\bf A Computational Approach to Measuring Vote Elasticity and Competitiveness}
\end{center}
  \medskip
} \fi

\bigskip
\begin{abstract}
The recent wave of attention to partisan gerrymandering has come with a push to refine or replace the laws that govern political redistricting around the country. A common element in several states' reform efforts has been the inclusion of \emph{competitiveness metrics}, or scores that evaluate a districting plan based on the extent to which district-level outcomes
are in play or are likely to be closely contested.

 In this paper, we examine several classes of competitiveness metrics motivated by 
 recent reform proposals and then evaluate their potential outcomes across large ensembles of districting plans at the Congressional and state Senate levels. 
 This is part of a growing literature using MCMC techniques from applied statistics to situate plans and criteria 
 in the context of valid redistricting alternatives.
 Our empirical analysis focuses on five states---Utah, Georgia, Wisconsin, Virginia, and Massachusetts---chosen to represent a range of partisan attributes.
 We highlight situation-specific difficulties in
creating good competitiveness metrics and show that optimizing competitiveness can produce unintended consequences on other partisan metrics.
These results demonstrate the importance of (1) avoiding writing detailed metric constraints into long-lasting constitutional reform and 
(2) carrying out careful mathematical modeling on real geo-electoral data in each redistricting cycle.
 \end{abstract}

\noindent%
{\it Keywords:}  Redistricting, Gerrymandering, Markov Chains, Competitiveness
\vfill

\newpage
\tableofcontents

\newpage
\spacingset{1.45} 
\section{Introduction}
\label{sec:intro}

In 2018 alone, five states saw redistricting reform measures passed by voters at the  ballot box, signaling a larger movement to revisit redistricting rules in state laws and constitutions in the run-up to the 2020 Census.
A common feature in this surge of reform efforts is redoubled scrutiny of the rules, priorities,
and criteria for new districting plans. In addition to reconsidering the treatment of traditional districting principles like population balance, compactness, and preservation of municipal boundaries, 
most of these state reforms explicitly address the use of partisan data and metrics in the redistricting process. 

One reform approach in the recent trend has been to bar mapmakers from partisan considerations. Several states broadly
prohibit favoring political parties or incumbents, while others go further and restrict mapmakers from directly using 
 any partisan data while drawing the lines. 
At the opposite extreme, other reform approaches mandate that certain types of partisan symmetry or fairness metrics should 
be important decision criteria, effectively requiring that partisan data be taken into account during map construction.
Competitiveness rules fall into the second category.\footnote{Some reform measures explicitly provide for a firewall between partisan-blind and partisan-aware
parts of the process.  For instance, Utah's successful ballot initiative in 2018 requires the maps to be drawn without considering partisan data but requires the expected partisan behavior to be measured before a plan can be approved \cite{UT}. Arizona requires similar behavior from its redistricting commission; the relevant portion of the redistricting law says that ``Party registration and voting history data shall be excluded from the initial phase of the mapping process but may be used to test maps for compliance...'' \cite{AZ}. In these cases, the process is designed to prevent mapmakers from directly optimizing for the metrics and instead sets up a failsafe against intended or unintended partisan skew.} 

This new attention to redistricting criteria is taking place in the context of an explosion of data and computational power in  
redistricting.  One consequence of this new era of computational redistricting is that methods for locking in partisan advantage---and 
data proxies that make this possible without explicitly optimizing for election outcomes---create major opportunities for abuse.  
At the same time, computationally sophisticated ensemble techniques are improving dramatically at detecting partisan outliers.  In this paper 
we devote our attention to a second, complementary application of algorithmically-generated ensembles of plans:  we use them to study the effects
of enacting different sets of rules and priorities.  In particular, we will develop a point of view
that ensembles can reveal some districting tasks to be more ``elastic" than others:
control of line-drawing always confers some control of outcomes, but some situations prove to
be more manipulable than others. 

Our main focus here is on rules favoring electoral competitiveness.
Competitiveness is frequently regarded as a healthy attribute of elections that promotes voter engagement, responsiveness, and accountability of representatives, among other benefits.\footnote{For instance, consider this language in Colorado's recent redistricting reform legislation: ``Competitive elections for members of the United States House of Representatives provide voters with a meaningful choice among candidates, promote a healthy democracy, help ensure that constituents receive fair and effective representation, and contribute to the political well-being of key communities of interest and political subdivisions'' \cite{CO}.} 
During an election, competitiveness of a race might be measured by close polling or by candidate spending;
after the fact, it is often assessed 
 by a close outcome.  It is much harder to capture the 
competitiveness level of districting plans at earlier stages, since they are created before there are outcomes, polling, or even a choice of candidates for the coming cycle.  Accordingly,  
 there is no agreed-upon means of measuring competitiveness,
 and states have operationalized it in different ways
when they have defined it at all.  
It is far from clear that a high competitiveness score can be achieved by a plan without other partisan consequences or unintended costs to other districting criteria, or that it will bring about outcomes that increase legitimacy and public trust.  
With these questions in mind, in this paper we develop statistical techniques to 
probe the consequences of various ways to quantify competitiveness of districting plans.

The primary reason that it is difficult to evaluate the potential consequences of altering or reordering redistricting criteria 
is that each set of rules and constraints constructs a different universe of valid plans, and those 
spaces are vast and complex.
Since the set of valid plans typically is too large to enumerate, 
Markov chain Monte Carlo (MCMC) methods are 
standard techniques in applied mathematics and statistics for sampling in these situations, used to gather representative samples from the space of interest.
Markov chain sampling is currently used by several groups of researchers to evaluate the partisan nature of proposed or enacted districting plans in court challenges \cite{hdsr, chikina_practical_2019,chikina_assessing_2017,redist, herschlag_quantifying_2018, herschlag_evaluating_2017}.  These methods work by constructing large ensembles of plans that satisfy the rules set forth in each state and comparing the statistics of enacted plans to aggregate statistics of the distribution. 
These techniques, however, require the choice of mathematical specifications of the legal constraints, 
which are rarely written with enough precision to determine an unambiguous formulation. 
Understanding the interactions between various districting criteria is fundamental for drawing meaningful quantitative conclusions from ensemble analysis, 
and we hope that this study provides an invitation for more research in this direction.

\subsection{Contributions}

We analyze several variants of competitiveness metrics and compare the effects of prioritizing them in different ways.
We use both {\em neutral ensembles} of alternative districting plans (i.e., generated without considering partisan data) and {\em heuristically-optimized} plans that seek to extremize the metrics for Congressional and state Senate plans in five states.\footnote{The only reason not to consider state House plans is that House districts tend to be too small to be made out of whole
precincts while maintaining tolerable population deviation.  Precincts are the smallest level at which we have authoritatively
accurate election outcomes, which are a crucial element of this style of analysis.  Vote results can be prorated to smaller units, 
but this is done at the cost of a potentially significant loss of accuracy.}
We aim this discussion at two primary audiences. The first is legislators and reform groups who are actively working to formulate new redistricting rules, particularly those considering imposing competitiveness criteria. 
The second audience is researchers and quantitative analysts who are interested in assessing how well proposed plans uphold the stated
criteria from which they were constructed.
A message that emerges for  both groups is the major impact of seemingly innocuous modeling choices against the 
idiosyncratic political geography of individual states.  We argue that no
  universal metric of competitiveness  should be adopted 
across the country, but rather that there is a fresh choice to be made in each circumstance.  On the other hand, we find that certain metrics are unreasonable and should \emph{not} be adopted.

For instance, packing and cracking are the tools of the trade for gerrymandering, so it is notable that in certain circumstances, a competitiveness metric can approve or even incentivize packed and cracked plans.   
For instance, in a state
with 68-32 voting for Party A, a plan that arranges A voters into some districts at 48\% share and compensates that with other districts at 88\% share could be textbook vote dilution while looking excellent under a metric that 
merely counts nearly-balanced districts.  If an arrangement like that were the only or the predominant way to satisfy
a competitiveness priority, then this would be important to know in advance of enactment.  A principal contribution
of our ensemble analysis is that it escapes the tendency to treat votes as though they are geographically unbound and fluidly rearrangeable,
instead studying the actual elasticity of the vote by comparing to alternative geographic partitions.

In our treatment, a general discussion of the reasons for preferring competitive districts in terms of broad alignment with societal and democratic goals will remain in the background. These normative questions are extremely important, but they are beyond the scope of the present paper and outside the reach of ensemble techniques, except to the extent that these techniques clarify the desirable and undesirable consequences of the norms.

\subsection{Related work}

Approaches to competitiveness in the political science literature vary widely.  
Although most recent reform efforts have focused on making more competitive districts, arguments have been made for instead packing districts with members of a single party to the extent practicable \cite{brunell}. 
The opposite situation was considered in \cite{forgette}, where regression analysis was used to detect correlations between traditional districting criteria and competitive districts in the 1990 and 2000 cycles. They found that rules prioritizing population constraints as well as preservation of communities of interest and municipal boundaries were associated with more competitive districts. Additionally, their analysis helped expose the complexity of interactions between multiple redistricting criteria and the tradeoffs that are inherent in the process. Our work extends theirs by using ensembles to consider the potential impacts of directly incentivizing competitiveness, evaluating the properties of many alternative
districting plans rather than the currently-enacted maps. 
 Our analysis does not model the impacts of whether the lines are drawn by legislatures, courts, or commissions. Several groups of political scientists have studied the relationship between the identity of the line-drawers and the degree of competitiveness in subsequent elections \cite{carson, cottrill, grainger,henderson, miller}, with mixed conclusions. 

A recent article by Cottrell \cite{cottrell} presents an ensemble analysis to evaluate the impact of gerrymandering on the level of political competition around the country, concluding that gerrymandering has only a small cost in competitiveness
because neutral processes also produce safe seats at rates similar to what can now be observed.
We refer the reader to his references 
for a broad discussion of the arguments around competitive standards in the political science literature.  
This differs from our approach both in goals and methods, but we view our problem domain as complementary to his. 
In terms of goals, Cottrell seeks evidence of gerrymandering, while we 
study the impact of rules that favor competitive outcomes.  
That means he is comparing neutral ensembles to enacted plans, and we are focused on comparing neutral ensembles to methods for favoring competitive outcomes. Another difference is that our analysis uses a newly-developed Markov chain technique with strong evidence of sampling effectiveness.\footnote{With the recombination technique used here, collecting 100,000 plans from a Markov chain
is shown to be enough in many circumstances to obtain a consistent distribution of sample statistics, regardless of starting point. See \cite{hdsr} and its references for details.}


\subsection{Review of competitiveness language in legal criteria}

Our research is motivated by several recent examples of redistricting rules introduced at the state level to promote
competitive districts. 

\subsubsection{Washington, New York, Arizona} 
Some states had competitiveness language on the books
before the recent reform wave, but not with 
quantitative details.  Washington requires its commission  to ``provide fair and effective representation and to encourage electoral competition'' and  New York's rules require that ``districts not be drawn to discourage
competition" \cite{WA,NY}.

Arizona's instructions to its commission  include the following language: ``To the extent practicable, competitive districts should be favored where to do so would create no significant detriment to the other goals'' \cite{AZ}.  
In practice, out of 9 districts, this text seems to 
suggest a conspicuous effort to craft 3 competitive
districts (53-47 outcomes or closer in 2012), leaving 4 safe Republican districts and 2 safe Democratic districts
(all 60-40 or farther).

\subsubsection{Colorado}
In 2018, Colorado voters approved a ballot initiative that included a requirement that plans ``maximize the number of politically competitive districts,''   where competitive is defined as ``having a reasonable potential for the party affiliation of the district's representative to change at least once between federal decennial censuses'' \cite{CO}. 
A literal reading of this language suggests that any district in which both parties have at least a 13\% projected chance of winning might match the Colorado law, since $(1-0.13)^5 \approx 0.5$. While this interpretation makes several potentially unrealistic
assumptions (including that a point prediction would carry through over the full cycle and that there is no impact of incumbency on the elections), the text of the legislation does not provide implementation details.\footnote{This rule  compares interestingly to the measure of competitiveness used by the 538 Atlas of Redistricting \cite{538}.  The Atlas project defined districts in the range from D+5 to R+5 as competitive, using the CPVI metric described in \S\ref{sec:band}.
Under their vote modeling, this corresponds to  projected  probabilities of approximately $18\% - 82\%$ of electing a candidate from a given party.} 

\subsubsection{Missouri}
Missouri's 2018 constitutional amendment, passed through a voter referendum known as Clean Missouri, defines competitiveness in terms of a partisan metric called the {\em efficiency gap}, or $EG$ \cite{MO}.  It is one of the more
detailed descriptions of a competitiveness metric in the recent reform 
measures, requiring the state demographer to create a 
blended voting index by computing the turnout-weighted averages of the Presidential, Senatorial, and Gubernatorial elections that occurred during the previous census cycle.
This pattern of votes is then modified by applying a uniform 
partisan swing of $\{-5,-4,\dots,+4,+5\}$ and recomputing
the efficiency gap against each modified vote dataset.
Proposed districting plans are to be constructed so that each of the corresponding $EG$  values is ``as close to zero as practicable"---we call this {\em $EG$/swing zeroing}. 
We will show in \S\ref{sec:eg} that this definition of competitiveness
promotes drastically different qualitative properties of districting plans than the other metrics, and if taken 
literally it actually puts an {\em upper limit} on the number of districts that are allowed to fall in the 45-55\% range for its voting index.   

Missouri's reform is also noteworthy in that it places a higher priority on compliance with the priority on competitiveness than on contiguity or compactness of the plan.

\subsubsection{New Jersey}
 Finally, in 2019, New Jersey's legislature debated a constitutional amendment  that defined a competitive district to be one where the expected voting outcome is within 5\% of the statewide average over the previous census cycle \cite{NJ}. A plan was defined to be sufficiently competitive if $25\%$ of its districts were competitive. 
The bill was widely pilloried (and ultimately defeated)
because in a state whose recent voting trends
$\approx 57\%$ Democratic, a district must have a
Democratic majority to count as competitive under
this language.

\bigskip

These examples and others demonstrate a range of efforts intended to quantify and promote ``competitiveness'' in districting plan design. 
Our analysis below defines  mathematical formalizations in keeping with the legal language reviewed above 
to study potential impacts of implementation.

\section{Competitiveness metrics}
We consider several types of competitiveness metrics in this paper, motivated by the recently proposed legislation in several states discussed above.

  \subsection{Plans and votes}
  \label{sec:delta}
  
  Characterizing a districting plan or individual district as competitive necessarily requires some assumptions about the underlying voter behavior.  
That is, if we denote a districting plan by $\D$ and
a vote pattern by $\Delta$, then all scores $C$ of competitiveness are really functions of both:  
$C=C(\D,\Delta)$.
The choice of $\Delta$ that allows the most robust assessment of
a Congressional or legislative districting plan is a matter of some debate.  Some political scientists insist that it is important to use only {\em endogenous} data, or vote patterns from the same type of election, but the partisan preference information in these patterns is confounded  by other variables such as 
incumbency, uncontested races, and particular candidate 
characteristics.  Other scholars prefer the use of statewide {\em exogenous} vote patterns, but then must make a case about the choice of which election or elections to use to create an informative voting index.  
See for instance \cite{best_authors_2017,best_considering_2017,mcghee_rejoinder_2017} for a paper and 
series of responses on this issue.
In all cases, care is required in terms of predictive claims about future voting.  We discuss the choice of $\Delta$ in more detail in \S\ref{sec:metrics}.  

Several states that require partisan analyses of proposed plans have addressed  the choice of election data  in their legislation. As noted above, Missouri specifies the exact election data that must be used to evaluate partisan performance \cite{MO}, while Colorado simply requires \cite{CO} that ``[c]ompetitiveness may be measured by factors such as a proposed district's past election results, a proposed district's political party registration data, and evidence-based analyses of proposed districts.'' New Jersey's proposed amendment \cite{NJ}, had it passed, also would have specified exactly which data was to be considered. As with many of the other variables in redistricting legislation, there does not appear to be a consensus around the optimal modeling choice. 

The results and visualizations reported below are based on a fixed vote pattern:  we 
use $\Delta=$Pres16, i.e., the precinct-level election
returns from the 2016 Presidential election, restricted only to the Democratic and Republican votes cast.  
No aspect of our method hinges on this choice, and it can be repeated for any other voting 
pattern in place of $\Delta$.

  \subsection{Evenness and typicality}
  
Perhaps the most intuitive measure of competitiveness for a district with respect to a vote pattern is the difference between the two-way vote share and 50\%.  
This metric is implicit in Arizona's and Colorado's legal language. Up to scale, this measures the number of votes needed to change the outcome in the district. We might consider this a measurement of {\em evenness}, and it aligns well
with standard ideas about competitiveness.

The Cook Partisan Voting Index (CPVI) is a widely used metric
that is constructed by comparing the major-party vote share in each geographical unit---precinct, district, etc.---to the nationwide average across the two most recent Presidential elections \cite{cvpi}. Frequently, a district is described as competitive if its CPVI value is between D+5 and R+5, meaning that the district's vote is within 5\% of the nationwide average. Upon inspection, most observers would agree that in the case that the  average is skewed 
away from 50-50, this is not measuring competitiveness but rather {\em typicality}.
Nonetheless, CPVI is reported by the Cook Political Report after each election as a measure of competitiveness across the nation; it also was used by 538's Atlas of Redistricting to count  competitive districts \cite{538}.
New Jersey's proposed measure described above was based on an idea similar to CPVI, except that districts are compared to the statewide instead of nationwide average.   Below, we will denote the statewide share of Democratic voters with $\Dbar$.

To measure the effects of quantifying and promoting competitiveness, we compute and compare metrics of 
evenness and typicality below with a generalized formulation that we call vote-band metrics.

  \subsection{Vote-band metrics}
  \label{sec:band}

We start by defining a district to be in the  {\bf $(y,z)$ band} with respect to voting pattern $\Delta$, or simply to be a $(y,z)$ district,  if its Democratic vote share is within $y$ of a target  $z$.
By default we will let $y,z$ be denominated in percentage points, so that for instance a district is in the $(5,50)$ band with respect to a voting pattern $\Delta$ if its vote share is between $45\%$ and $55\%$.

Aggregating this behavior over the districts in a plan, we define $(\D,\Delta)$ to be {\bf $(x,y,z)$ compressed} 
if at least $x$ share of the districts are in the $(y,z)$ band.    By varying the parameters $x$, $y$, and $z$, we can understand
the elasticity of a vote pattern $\Delta$ in terms of its divisibility into districts at a certain scale.  
If voters of one group are very clustered, there might be a great deal of control in the hands 
of the districters by manipulating how the lines cross the clusters.  In case of extreme dispersal,
there may be little impact to moving the lines.  
And in between those extremes, the spatial distribution of votes and the units of data discretization combine to create effects on the range of possible outputs.
Examining whether it is possible to compress 
district-level vote shares into narrow bands of outcomes is one way to examine how much 
control the districters have to create districting plans with  precise properties.

    
A legal standard that employs this type of metric might require that adopted plans  be 
$(x,y,z)$ compressed, where $x$, $y$, $z$, and the choice of $\Delta$ are set by statute.  This encompasses both many 
competitiveness rules (where the target is $z=50$) and typicality rules (where the target is $z=\Dbar$ for state typicality).  
We will consider how plans that satisfy compression properties relate to unconstrained plans, along with studying how varying the parameters affects expected outcomes as a function of political geography.
        
\begin{figure}
    \centering
\begin{tikzpicture}[xscale=1.6,yscale=2.2]    
\node at (0,10) {\includegraphics[width=160pt]{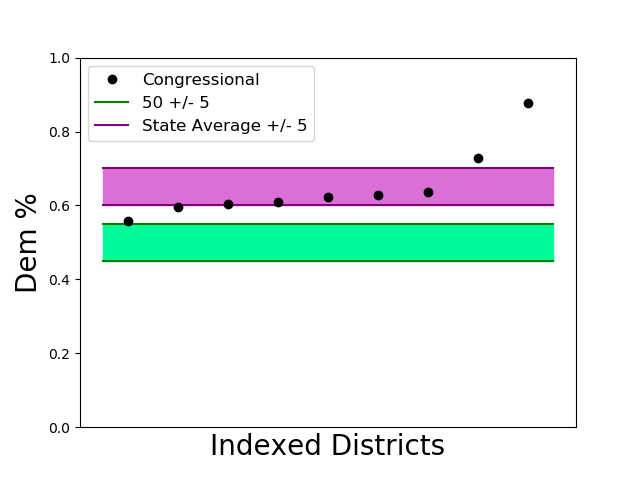}};    
\node at (0,8) {\includegraphics[width=160pt]{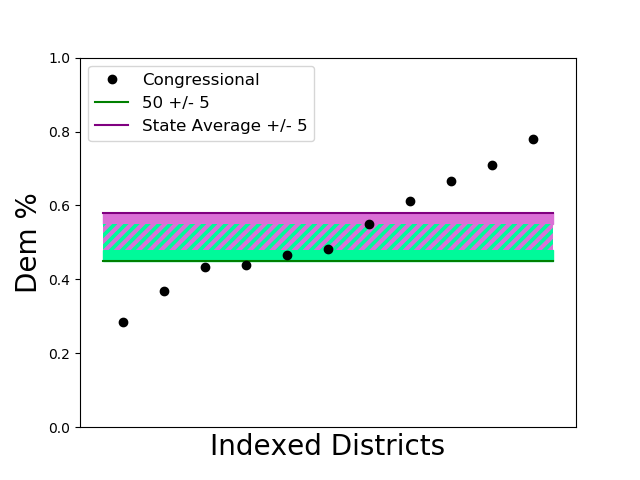}};    
\node at (0,6) {\includegraphics[width=160pt]{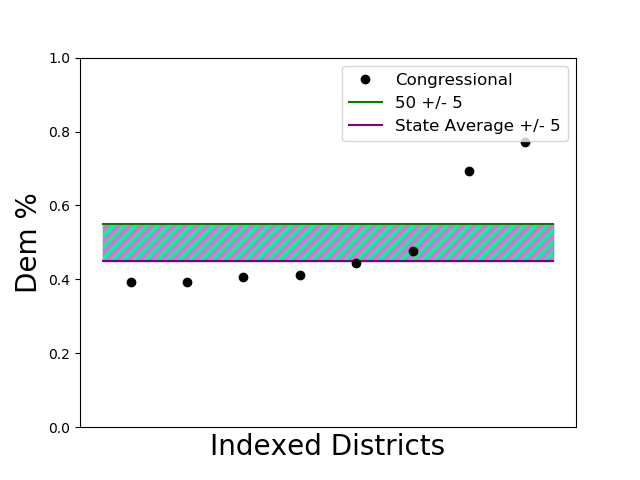}};    
\node at (0,4) {\includegraphics[width=160pt]{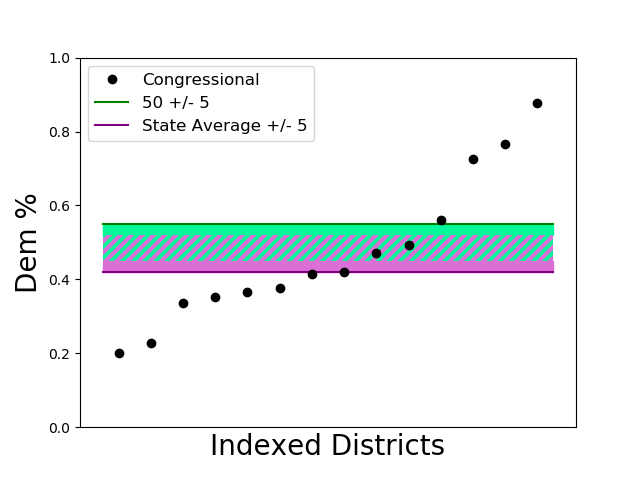}};    
\node at (0,2) {\includegraphics[width=160pt]{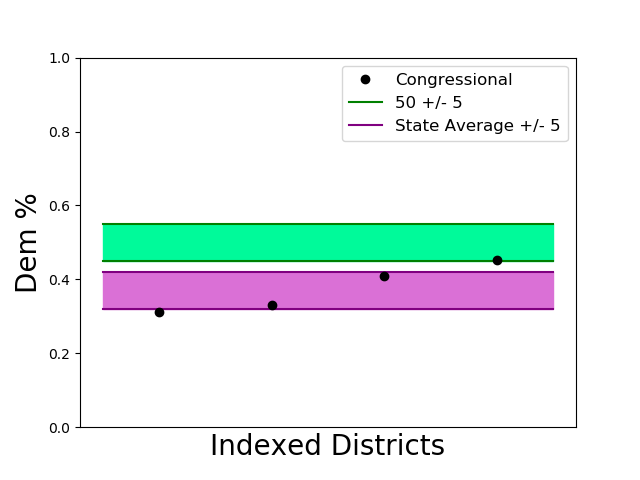}};  

\node at (2.5,10) {MA};
\node at (2.5,8) {VA};
\node at (2.5,6) {WI};
\node at (2.5,4) {GA};
\node at (2.5,2) {UT};

\node at (5,10) {\includegraphics[width=160pt]{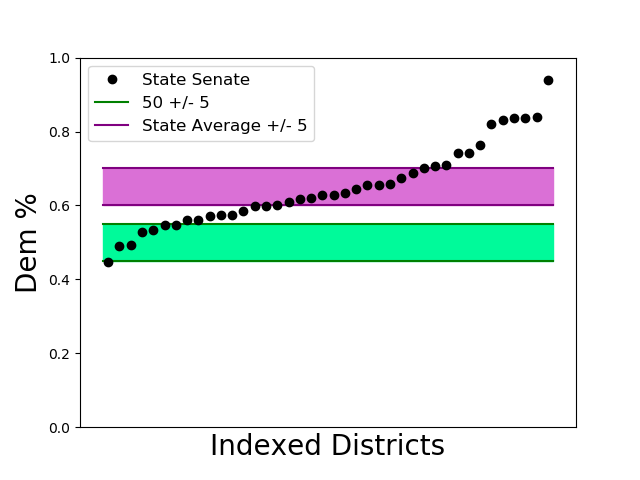}};    
\node at (5,8) {\includegraphics[width=160pt]{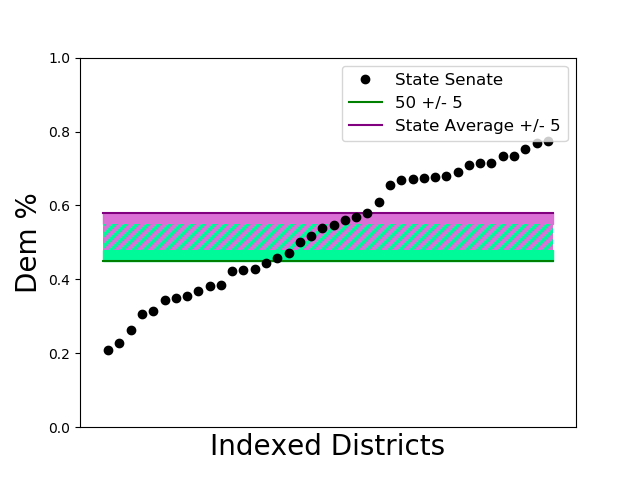}};    
\node at (5,6) {\includegraphics[width=160pt]{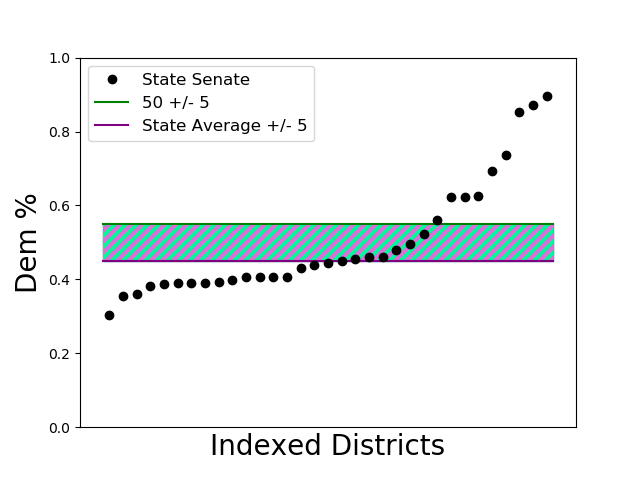}};    
\node at (5,4) {\includegraphics[width=160pt]{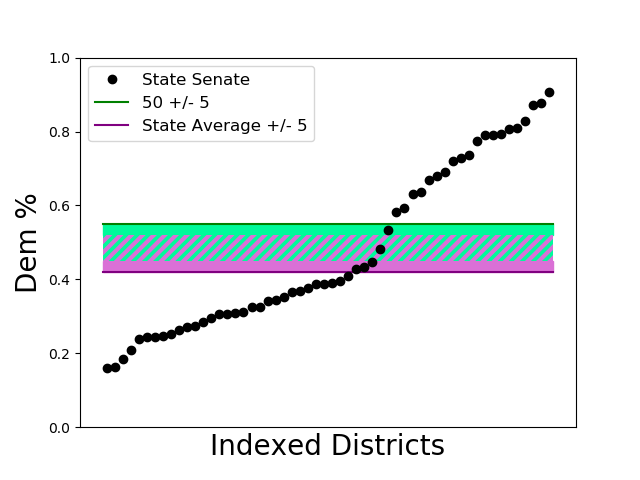}};    
\node at (5,2) {\includegraphics[width=160pt]{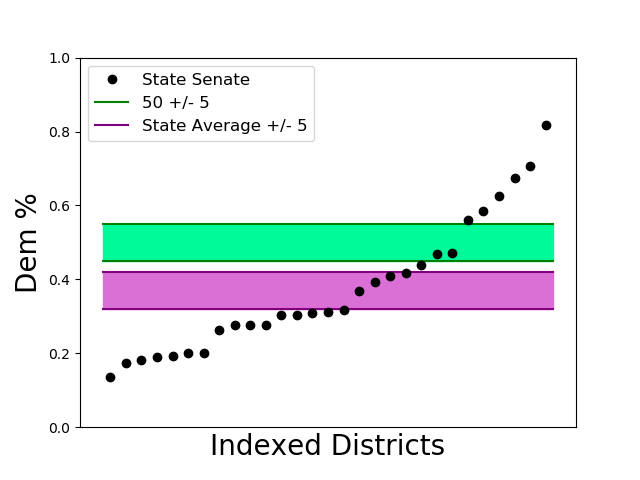}};    
\end{tikzpicture}    
\caption{Comparison of the enacted Congressional and state Senate  plans to $(5,50)$ bands (green, competitive) 
and $(5,\Dbar)$ bands (purple, state-typical). Here and in most plots below, the districts are ordered from smallest D share to highest to facilitate comparisons across plans.}
\label{fig:stateband}
\end{figure}
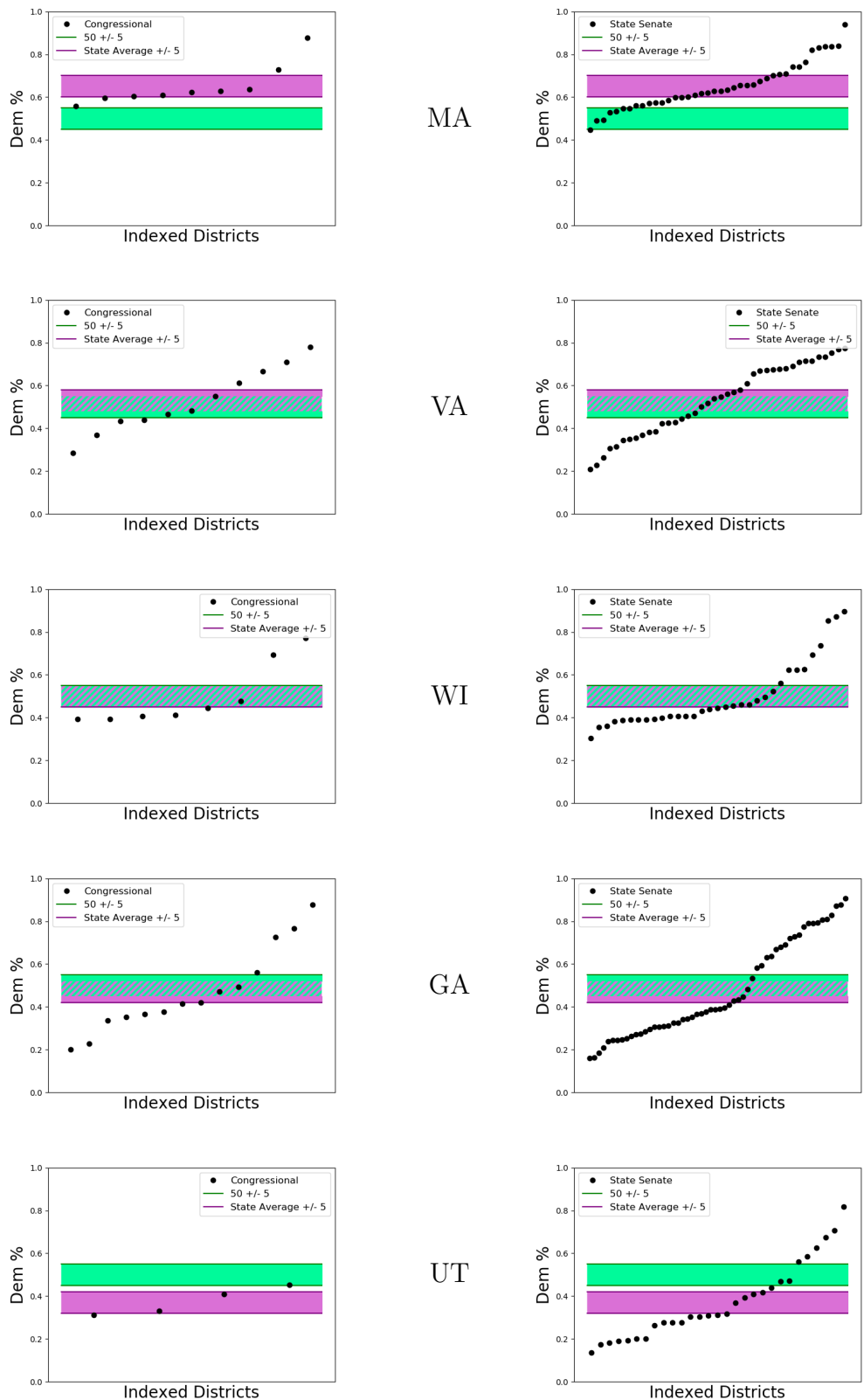

Figure \ref{fig:stateband} shows how the currently enacted Congressional  and state Senate plans in Utah, Georgia, Wisconsin, Virginia, and Massachusetts relate to competitiveness and state-typicality bands at $y=5$,
using 2016 Presidential vote data.  Table \ref{tab:avg50} records the number of districts in those plans that fall into each band. Although none of these states had requirements governing competitiveness when the plans were drawn, they provide a starting example for considering the consequences of  vote-band rules.

\begin{table}[!h]
    \centering
    \begin{tabular}{|l|c|c|c|c|}
    \hline
         State&\# Cong &\# Sen & $\Dbar$ in  &2012-16 CVPI \\
         & Districts&Districts&Pres16& \\
         \hline
        Massachusetts&9&40& 64.7& D+12 \\
         \hline
         Virginia&11&40& 52.8& D+1\\
         \hline
         Wisconsin&8&33& 49.6& Even\\
         \hline         
         Georgia&14&56& 47.3& R+5\\
         \hline
         Utah&4&29& 37.6& R+20\\
         \hline
    \end{tabular}
    \caption{The five states considered in our study, chosen for a wide range of partisan tilts. The Democratic share $D_0$ is the two-party share in Pres16, so that for instance the sizeable third party vote in Utah is not reflected.}
    \label{tab:statebands}
\end{table}
    
In most states, several districts lie outside both bands. 
Given fixed $(y,z)$, there is no guarantee that it is even possible to construct a plan with a large number of $(y,z)$ districts---even if $z$ is near the statewide average---while adhering to reasonable compactness and boundary preservation norms. 
In Virginia, the enacted plan almost fails to have any districts that are competitive under both definitions, even though there is a large overlap between the bands.

One straightforward observation is that there must be some districts at least as far from 50\% as the state average. Additionally, as shown in \cite{MGGGMA}, discretization of geographical units can play a large role in determining the range of possible outcomes, since the voting population is not infinitely divisible.

\begin{figure}
\centering
\begin{tikzpicture}[xscale=1.6,yscale=2.2]     
\node at (0,10) {\includegraphics[width=160pt]{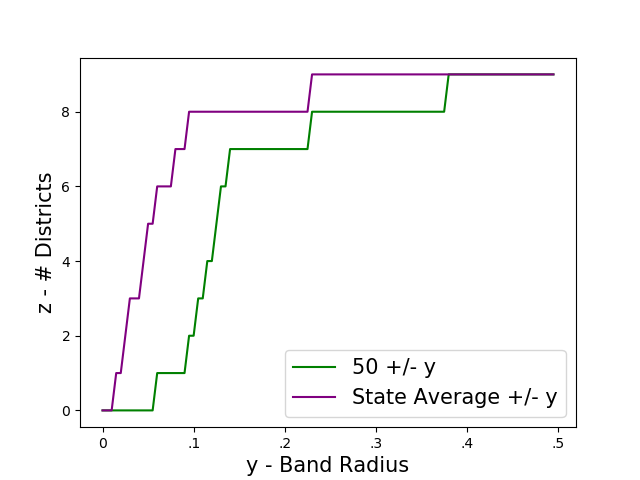}};    
\node at (0,8)  {\includegraphics[width=160pt]{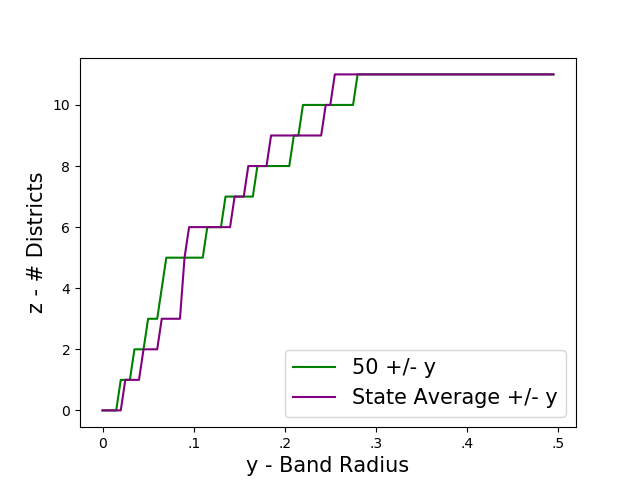}};    
\node at (0,6)  {\includegraphics[width=160pt]{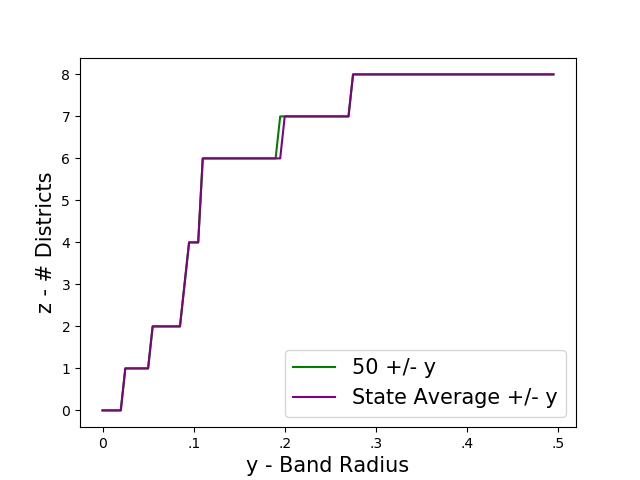}};    
\node at (0,4)  {\includegraphics[width=160pt]{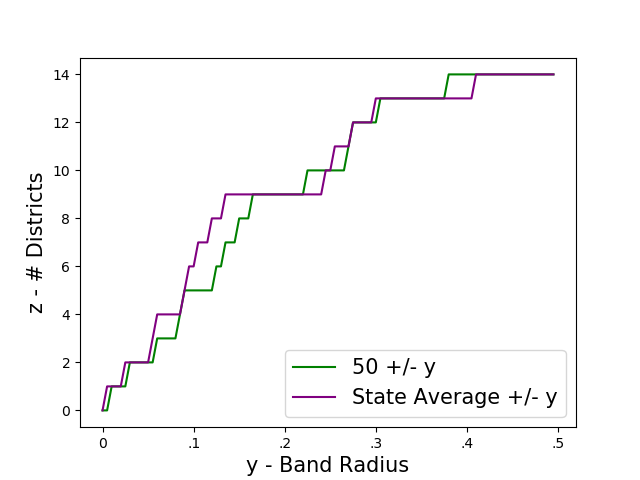}};    
\node at (0,2)  {\includegraphics[width=160pt]{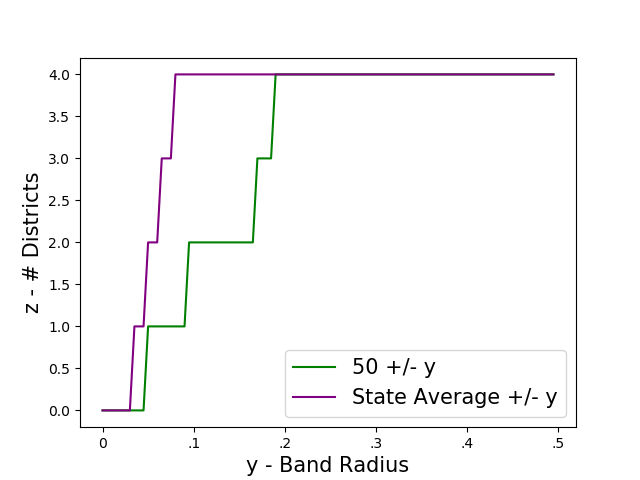}};    

\node at (2.5,10) {MA};
\node at (2.5,8) {VA};
\node at (2.5,6) {WI};
\node at (2.5,4) {GA};
\node at (2.5,2) {UT};
    
\node at (5,10) {\includegraphics[width=160pt]{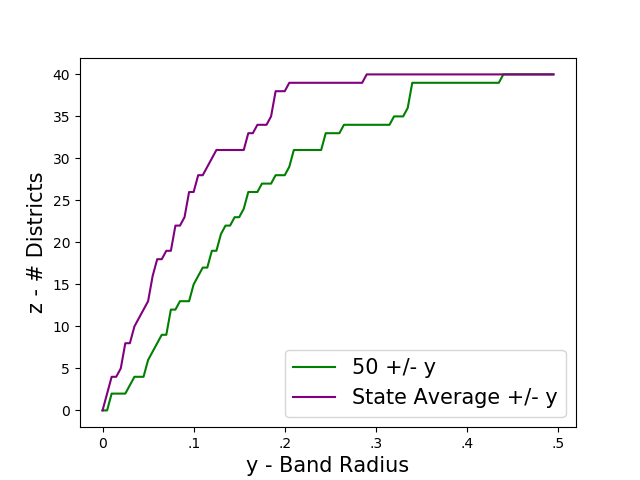}};    
\node at (5,8) {\includegraphics[width=160pt]{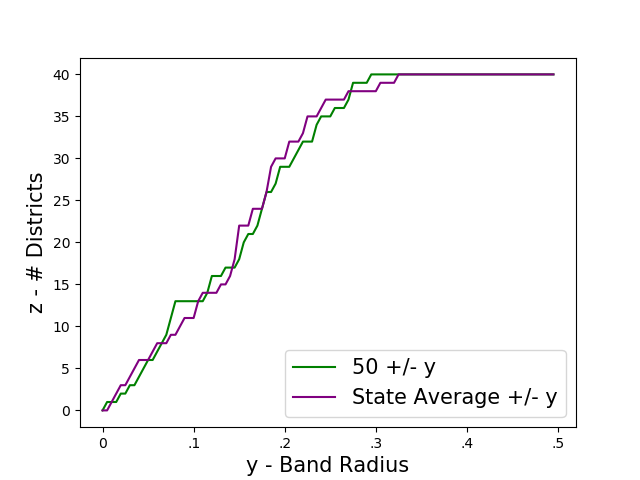}};    
\node at (5,6) {\includegraphics[width=160pt]{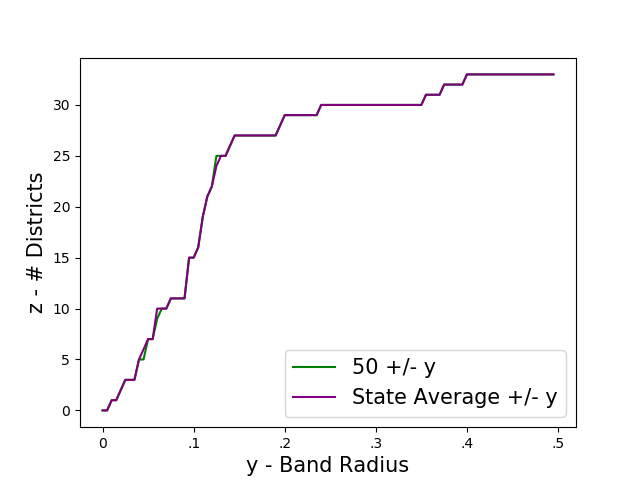}};    
\node at (5,4) {\includegraphics[width=160pt]{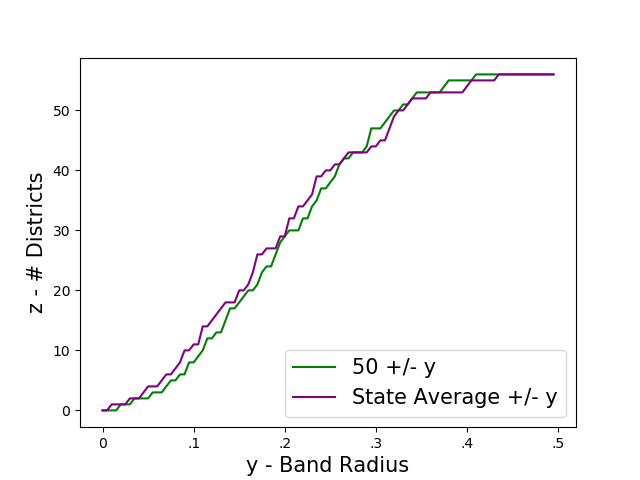}};    
\node at (5,2) {\includegraphics[width=160pt]{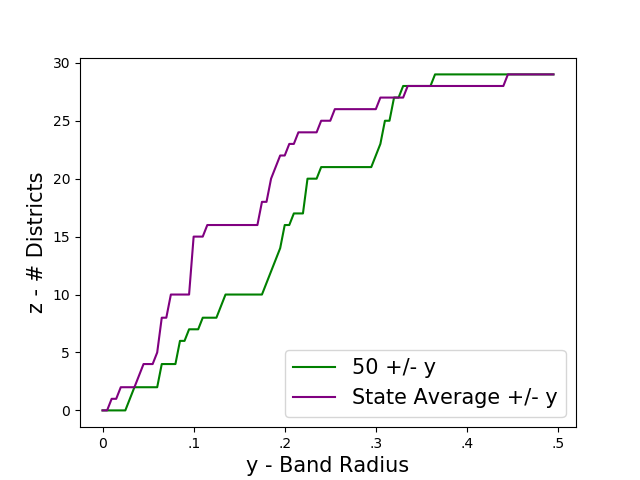}};    
\end{tikzpicture}    
\caption{Number of districts in the enacted Congressional and state Senate  plans that are $(y,50)$ (green, competitive) 
and $(y,\Dbar)$ (purple, state-typical), as $y$ varies.
For instance, all four of Utah's congressional districts are within 10\% of the state average.}
    \label{fig:slidingband}
\end{figure}
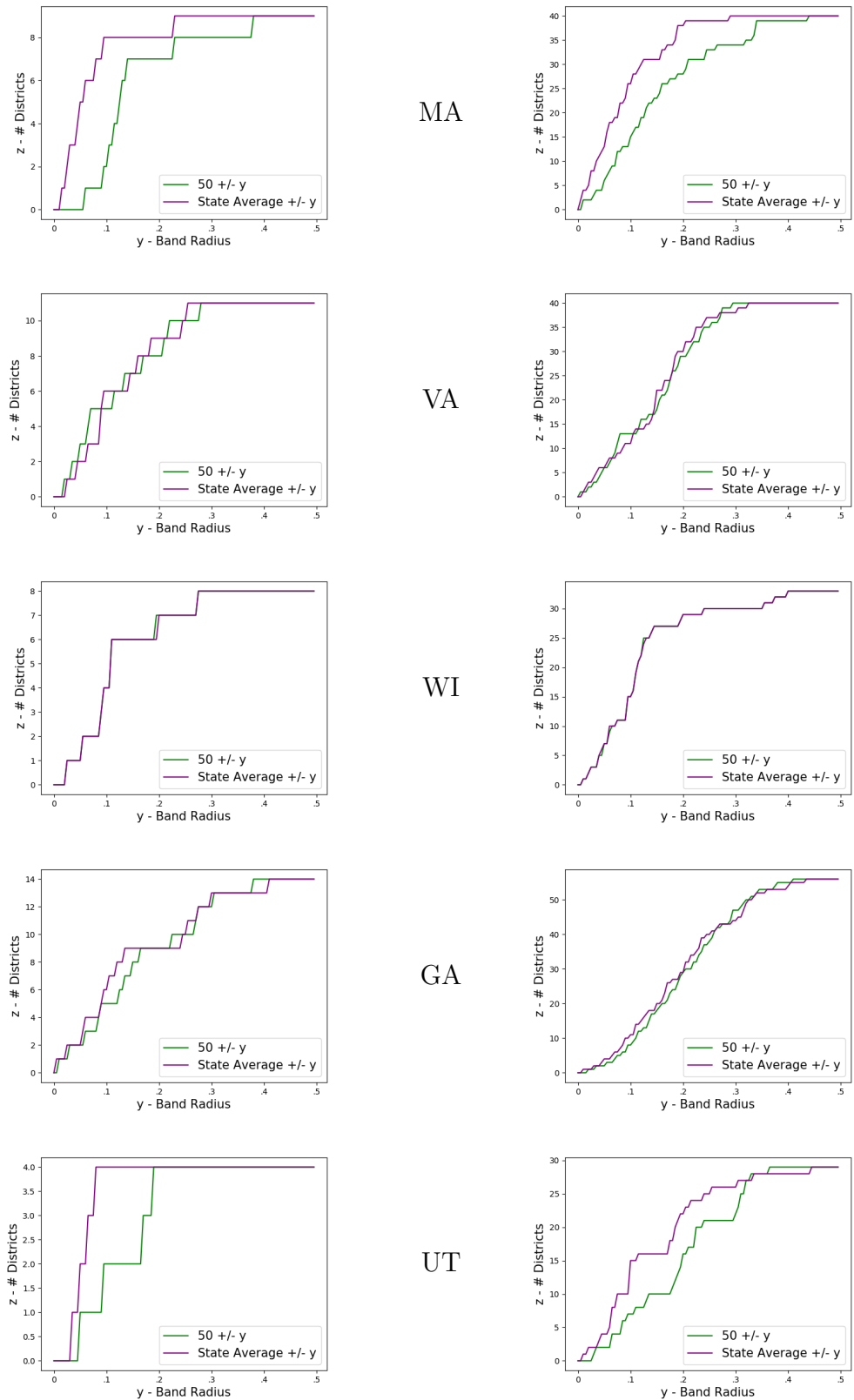

Given a fixed vote share target $z$, it is natural to ask how the number of $(y,z)$ districts varies as a function of 
the band width parameter $y$.
  Figure \ref{fig:slidingband} shows the number of districts in the enacted plans that lie in competitive and state-typical bands as the width is increased.  For most of the current plans, $y$ values of approximately 15-20\% would be needed to make half of the districts competitive; that is, about half the districts have outcomes even more skewed than 70-30. 
Also, as one would expect, it is harder to make even districts in 
the states that are themselves more partisan-tilted.  Comparing Utah and Massachusetts to the other states shows 
the extent of this effect.
    
In our experiments, we will explore the interaction of vote-band metrics with political geography. Although this formulation is simple, we will see that the impacts of enforcing a choice of parameters $x$, $y$, and $z$ do not fall uniformly over our 
test states.
  
 \subsection{$EG$/swing zeroing}
 \label{sec:eg}\label{sec:EG-swing}
As discussed above,
Missouri has adopted a definition of competitiveness for redistricting that requires the State 
Demographer to compute a metric called the \emph{efficiency gap} (described below) for any proposed districting plan, with respect to a particular synthetic vote pattern created by a weighted combination of recent election outcomes. They are also required to consider ten alterations of that vote pattern, obtained by uniformly swinging the district vote shares by $\{\pm1, \pm2, \pm3, \pm4, \pm5\}$ and computing the corresponding efficiency gap values. 
A literal reading of the constitutional text requires that each of those eleven efficiency gap values should be made as small as practicable. 
Due to the close relationship  between efficiency gap and seat outcomes, many authors have noted that $EG\approx 0$  prescribes a ``winner's bonus" with slope two in the votes to seats conversion \cite{eg,ev,formula}.\footnote{That is, having $EG\approx 0$ requires having a tight relationship
between overall vote share and overall seat share, with each additional percentage point of votes
calling for {\em two} additional percentage points of seats.  Since seats must turn out in 
a particular way to satisfy the $EG$ test, there is a completely different reason that it is at odds with competitiveness:  competitive seats introduce uncertainty and the potential for an unpredictable $EG$ score \cite{formula}.} 
Here, we show that this constraint is directly at odds with vote-band competitiveness.

Let $\Dbar$ represent the statewide percentage  of Democratic voters  and $S$ represent the number of seats won by the Democratic party, out of $k$ total seats. Then, as derived by Veomett in \cite{ev}, the efficiency gap is 
$$EG(\D,\Delta)=2\left(\frac\Dbar{100}\right)-\frac Sk - \frac 12 + N(\D,\Delta),$$
where $N(\D,\Delta)$ is a noise term depending on
$S,k,$ and the average turnout in districts won by each party, which is zero in the case of equal turnout.\footnote{Indeed,  publications by Stephanopoulos--McGhee sometimes use this full definition 
of $EG$ and sometimes use the ``simplified formula" without the $N$ term.  Converting from Veomett's notation to ours and letting $s=S/k$ and $\rho$ be the ratio of average turnout in districts won by Party A to those won by Party B, we have 
$N=\frac{s(s-1)(1-\rho)}{s(1-\rho)+\rho}$. This vanishes in the equal-turnout case $\rho=1$.}
Noting that $\Dbar$ is fixed and neglecting the noise term, we must minimize this expression to satisfy the constitutional requirement,
which amounts to securing a number of Democratic seats $S$ as close as possible to $k(2\frac{\Dbar}{100}-\frac12)$. Repeating this procedure for each of $\{\Dbar-5, \Dbar-4, \ldots, \Dbar+4, \Dbar+5\}$ gives us a sequence of eleven prescribed seat values, $S_{-5}\leq S_{-4} \leq \cdots S_4\leq S_5$. 
But we further observe that prescribing how the seat values
change as you shift the votes 5 points up and down is 
precisely equivalent to prescribing the number and 
position of seats in the $(5,50)$ band.  

To show the potential impacts of this $EG$/swing 
zeroing rule, consider Missouri and Wisconsin.  With respect to Pres16, Missouri had roughly $\Dbar=40$.\footnote{Clean Missouri prescribes that the vote pattern be constructed from the 
last three races for Governor, Senate, and President.  At the time of writing, this amounts to a $\Dbar=50.3$, but when it is computed in 2021 it is likely to drop to the mid-40s.  Interestingly,
the number of competitive districts required by the rule is barely sensitive to the value of $D_0$.}
First we work out the consequences of the competitiveness rule on the 8-seat Congressional map, though we note that the Clean Missouri constitutional reform only applies to legislative districting.
Continuing to neglect turnout noise, we observe that
$EG(\D,\Delta+i)=2(\frac{\Dbar+i}{100})-\frac{S}{8}-\frac 12$, which prescribes
$S=2$ Democratic seats  
for shifts between $R+5$ and $R+0$ and by $S=3$ seats for $D+1$ through $D+5$.  This can {\em only} be satisfied
by having one district with a Democratic share of between
49 and 50 percent and no others between 45 and 55 percent.
If the Democratic shares by district were roughly $(22,22,22,22,22,49.5,80,80)$, the plan would earn perfect marks for ``competitiveness" by this $EG$-swing definition.  Having more districts close to even would rate strictly worse.  

With a bit more arithmetic, we can extract the pattern.  Noting that minimizing $EG$ amounts to solving $\frac Sk \approx 2V-\frac 12$,
we replace $V$ with its extreme values $\frac{D_0-5}{100}$ and $\frac{D_0+5}{100}$.  The difference between the largest and smallest number of required D seats is exactly 
the number required to fall in the 45-55 range.  
That is, accounting for rounding to the nearest integer, this rule prescribes that the number of $(5,50)$ seats is equal to $k/5$ or one-fifth of the available seats, plus or minus no more than one,
no matter whether $D_0$ is 35 or 40 or 45 or 50.\footnote{For values of $D_0$ more extreme than 30-70, the allotment for close seats actually goes down as the efficiency gap demands virtually all safe seats for the majority party.}
So, in Missouri's 34-seat Senate, the rule will require precisely 6-7 seats to fall between 45 and 55 percent, with at least 27 safe seats, and in the 163-seat House, it will require 32-33 close outcomes,
with at least  130 safe seats. 
 
In Wisconsin, which had roughly $\Dbar=49.6$, the $EG$-swing standard would require two Congressional districts between 45 and 55 percent and six districts outside that range. 
Again, this optimization promotes the creation of many safe districts and is indifferent to the possibility that those are actually made into landslide districts.
This effect is more dramatic in Utah ($\Dbar=37$, $k=4$), where adherence to this 
rule would require that Democrats are ahead in precisely one Congressional district out of four, for all of the swing values.  That is, a 
Utah plan complying with this definition of competitiveness would not be  allowed to have {\em any} districts within  45 and 55 percent.

This construction is in tension with the plain-English notion, in which a plan is said to be more competitive when more districts are close to even.  
In some states, demanding that one-fifth of seats (and no more) be competitive might be a reasonable goal and an improvement over the status quo; in other states, it will not be possible 
or will have major unintended consequences.  Considering that this definition also entails a specific number of safe seats for each side, it is not easily separated from a very prescriptive partisan
outcome rule and clearly calls for careful partisan tailoring.  
We will not continue to model more refined consequences of adopting this kind of metric below, but will focus on vote-band definitions.

\section{Data and methods for ensemble analysis}

We analyze the potential effects of enforcing a vote-band rule
as described in \S\ref{sec:band} in two sets of experiments, using data from five states. 
First, we generate a large neutral ensemble of districting plans for each state and filter the plans according to increasingly strict vote-band constraints, comparing the properties of the compressed plans to those of the full ensemble. Next, we use two heuristic optimization techniques to generate plans that score well on the given competitiveness measure and compare the statistics of the optimized plans to that of the full ensemble.

\subsection{State data}

In our study, we focus on Georgia, Massachusetts, Utah, Virginia, and Wisconsin, as exemplar states that display a wide range of  partisan leans (see Table \ref{tab:statebands}).  The CVPI baseline over this time period is approximately $51.5$\%, as the Democratic party candidate narrowly won the popular vote in both the 2012 and 2016 elections. 

Our methodology uses a discrete formulation of redistricting, viewing a plan as a graph partition for each given state. This is defined with respect to the {\em precinct dual graph}, which  has a node for each precinct and an edge between two nodes if the corresponding precincts are adjacent. (``Dual graph" is a mathematical term for a graph that records adjacency patterns.)
Given a fixed number of districts $k$, a districting plan is a partition of this graph into $k$ connected subgraphs.  Each node is decorated with the population of the corresponding precinct and its vote counts with respect to the election $\Delta$, so the population and partisan performance of each district can be computed by summing over the nodes.
This discrete approach has enabled a growing literature
on sampling techniques in court cases and in scientific
literature
\cite{cr,chikina_assessing_2017,hdsr,herschlag_quantifying_2018}. 

To construct the dual graph for each state, we use precinct
shapefiles with population and vote data joined.%
\footnote{The underlying geographic and partisan data for each state was originally obtained from the following sources.
 {\bf Georgia:}  Geography from the  Georgia General Assembly Legislative and Congressional Reapportionment Office \cite{GAdata}; partisan data from MIT Elections Data and Science Lab (MEDSL) \cite{mitedsl}.
  {\bf Utah:} Geography from the Utah Automated Geographic Reference Center \cite{UAGRC}; partisan data from MEDSL. 
{\bf Wisconsin:} All geographic and partisan data from the Wisconsin Legislative Technology Services Bureau \cite{WLTSB}.
 {\bf Virginia:} All geographic and partisan data from the Princeton Gerrymandering Project OpenPrecincts project \cite{openprecincts}.
{\bf Massachusetts:}  Geography from the Massachusetts Secretary of the Commonwealth and partisan data from the Massachusetts Secretary of the Commonwealth Elections Division \cite{msced}.
Staff of the Metric Geometry and Gerrymandering Group joined population data from the 2010 US Census to these datasets by aggregating up from census blocks,
then used geospatial tools to clean the geographical data.  The processed shapefiles and metadata are publicly available for \href{https://github.com/mggg-states}{download}  \cite{mggg-states}.}

\subsection{Partisan metrics}
\label{sec:metrics}

The choice of Pres16 as the vote pattern $\Delta$ for these runs makes it easier to compare results across states and is consonant with the CVPI approach and with several other political science analyses.  We note, however, that the resemblance between Presidential and  Congressional or Legislative voting patterns varies greatly from state to state and from year to year, and we do not feel that the lessons learned from this analysis hinge on there being a close match.

Nevertheless, we can observe that the number of districts with a Democratic majority in Pres16 matches the actual Congressional
2016 outcome in MA, GA, and UT (with 9/9, 4/14, and 0/4
Democratic-majority districts, respectively), while assigning
one D seat too many in Virginia and one too few in Wisconsin (where the enacted plan laid over Pres16 voting gives  5/11 and 2/8 Democratic-majority districts, respectively).
We have also carried out the  trials reported below with respect
to recent U.S. Senate and Gubernatorial elections where data was available; the findings are broadly similar, with the same big-picture messages about the relative
elasticity and the between-state variability of voting patterns.

To evaluate the partisan side effects of competitiveness rules on districting plans, we use three partisan metrics, again without taking a stance on which of these is the best or most useful.  Rather, we seek the presence of interrelations.
The first partisan metric is simply the number of districts in which the number of Democratic votes exceeds the number of Republican votes (discussed above in this section).
A second metric, defined in \S\ref{sec:EG-swing},
is the efficiency gap.  
Finally, a third popular partisan metric is the 
{\em mean-median ($MM$) score}, a measure of asymmetry that is defined by subtracting the mean Democratic vote share across the districts of a plan from the median of the same list of values.
Drawing on the work of Bernard Grofman, Gary King, and others, this metric is sometimes described as measuring how much one party could fall short of 
50\% of the statewide votes while still receiving 50\% of the seats \cite{mm1,mm2}.
Symmetry metrics are explicitly included in Utah's reform language from 2018 \cite{UT}.

\subsection{Ensemble generation}
\label{sec:methods}

We use  \href{https://github.com/mggg/gerrychain}{open-source  software \texttt{GerryChain}} \cite{gerrychain} both for generating the neutral ensembles and for optimizing for competitive districts. \texttt{GerryChain} uses a Markov chain to generate districting plans, beginning with an initial state and making iterative changes to the assignments of the nodes. Our ensembles are constructed using the {\sf ReCom} Markov chain procedure overviewed in \cite{hdsr}, which uses a spanning tree method to bipartition pairs of adjacent districts at each step.

For the neutral ensembles, we require the districts to be connected and population-balanced, imposing no further restrictions.\footnote{As described in \cite{hdsr}, 
{\sf ReCom} produces plans with compactness scores in range
of human-made plans without any need to impose additional
compactness constraints. Compactness is measured by a discrete
metric called {\em cut edges}, which counts how many pairs
of units were adjacent in the state but are assigned to different districts---these are the edges that would have to be cut to separate the whole graph into its district subgraphs.}
We constrain the population to within 2\% of ideal for Congressional districts and within 5\% for state Senate districts.\footnote{Most states balance Congressional districts to within one person, but 
2\% balanced plans can readily be tuned to tight balance by a skilled human mapmaker.  With 1 or 2\% deviation, we get efficiently moving Markov chains.  In other studies we have confirmed that 
loosening or tightening population balance does not have substantial impacts on partisan summary statistics \cite{VA-criteria}.} 
For each state, we take  100,000 {\sf ReCom} steps for the Congressional plans and 1,000,000 steps for state Senate plans, beginning at a random seed generated by a recursive tree-based partition;
plans with more districts will require more recombination moves to achieve a high-quality sample.
See \cite{hdsr} for a workflow describing 
standard heuristic tests of sample convergence, such as independence of seed and stability under increases in run length.

For the heuristic optimization ensembles, we use two types of hill-climbing proposals to find elevated numbers of $(5,50)$ districts. We call these methods {\sf Opt1} and {\sf Opt2}  and construct 100 plans using each.   We can take far fewer steps here than in constructing ensembles because we are not trying to sample, but only to find extreme examples.  
To generate each plan, we start with a number of unconstrained {\sf ReCom} steps (200 for Congress, 1000 for Senate) and then attempt 1,000,000 {\sf Flip} steps, each proposing to change the assignment of a single node on the boundary to move toward a more competitive plan.  The unconstrained steps serve to decrease correlation to the last map before attempting the greedy {\sf Flip} proposals.  
In the {\sf Opt1} run, 
proposed plans are rejected if they decrease the number of 
$(5,50)$ districts.  In the {\sf Opt2} run, a proposed plan is 
accepted if it decreases the sum of the distances between
the district Dem shares and the 45-55 percent band.  
Because {\sf Flip} steps tend towards poor compactness scores,
these steps are additionally constrained to have no more than twice 
as many cut edges as the partition after the {\sf ReCom} steps.

\section{Ensemble results}
\label{sec:results}
In the previous section, we describe how to generate large ensembles of compact, contiguous, population-balanced maps that do not take partisan data into account.
It is important to note how we propose to use these neutral ensembles for comparisons.  We do not propose that matching summary statistics to, say, the mean or median of a neutral ensemble is a normative goal in redistricting, and we do not view the neutral ensembles as strictly bounding the range of possibilities.  Instead, they provide a baseline range for any scores and metrics we might study,
constructed only by the basic rules of redistricting and the political and physical geography of the state.
 
The key takeaway is that baseline ranges for partisan metrics 
are sensitively dependent both on the formulation of the competitiveness metric and on the underlying political geography, which controls the elasticity of the vote.  For this reason, great care should be taken to evaluate potential impacts of these rules, using data representing the actual distribution of voters across the state,  before adoption.

\subsection{Partisan baselines}

\begin{figure}
    \centering
\begin{tikzpicture}[xscale=1.6,yscale=2.2]    
\node at (0,10) {\includegraphics[width=160pt]{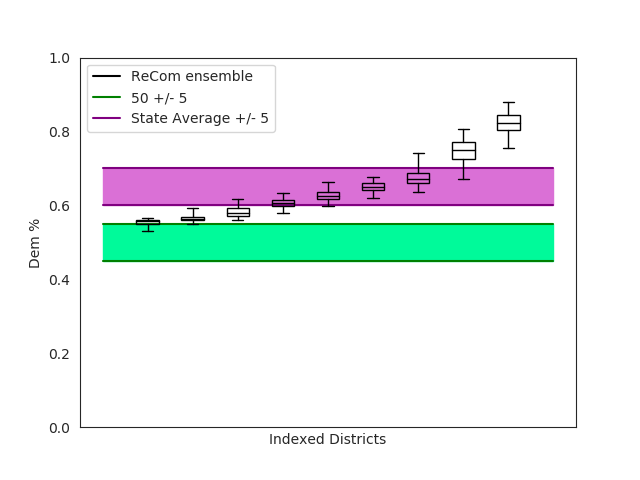}};    
\node at (0,8) {\includegraphics[width=160pt]{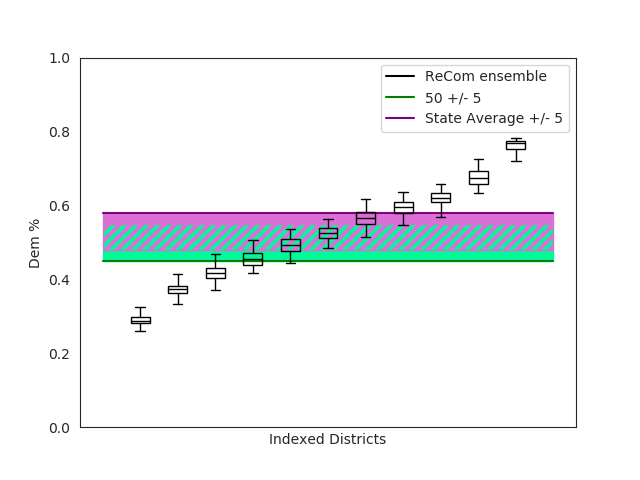}};    
\node at (0,6) {\includegraphics[width=160pt]{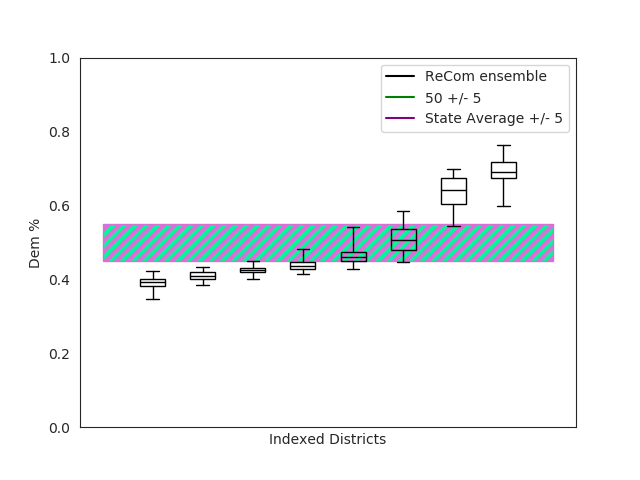}};    
\node at (0,4) {\includegraphics[width=160pt]{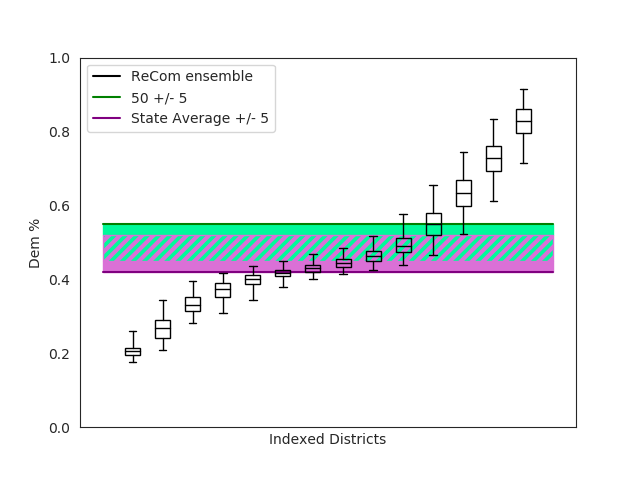}};    
\node at (0,2) {\includegraphics[width=160pt]{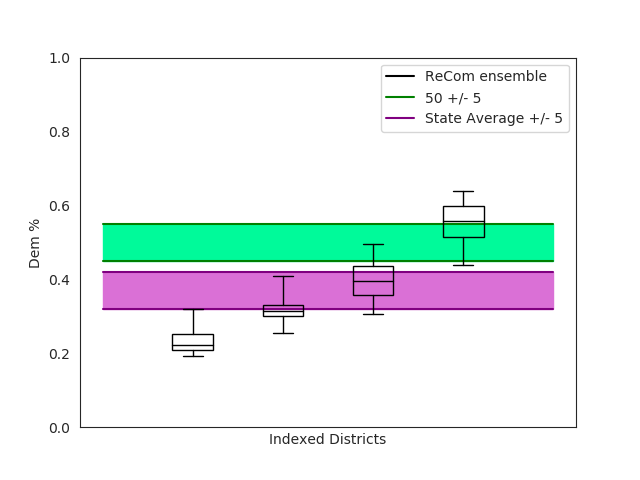}};  

\node at (2.5,10) {MA};
\node at (2.5,8) {VA};
\node at (2.5,6) {WI};
\node at (2.5,4) {GA};
\node at (2.5,2) {UT};

\node at (5,10) {\includegraphics[width=160pt]{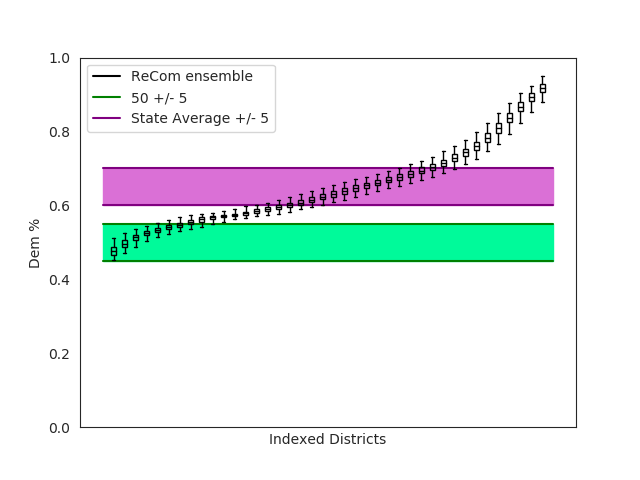}};    
\node at (5,8) {\includegraphics[width=160pt]{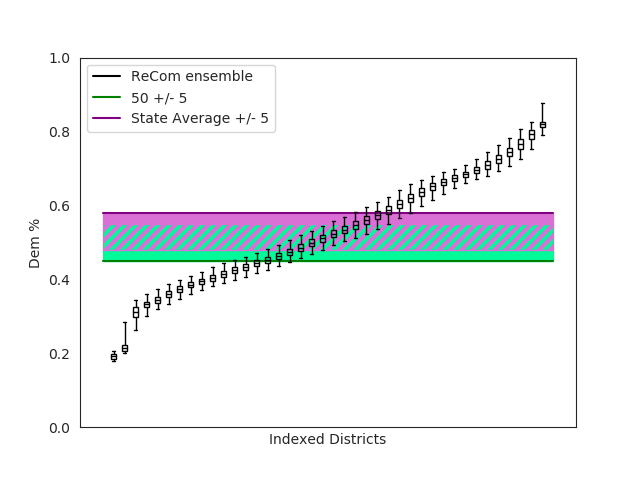}};    
\node at (5,6) {\includegraphics[width=160pt]{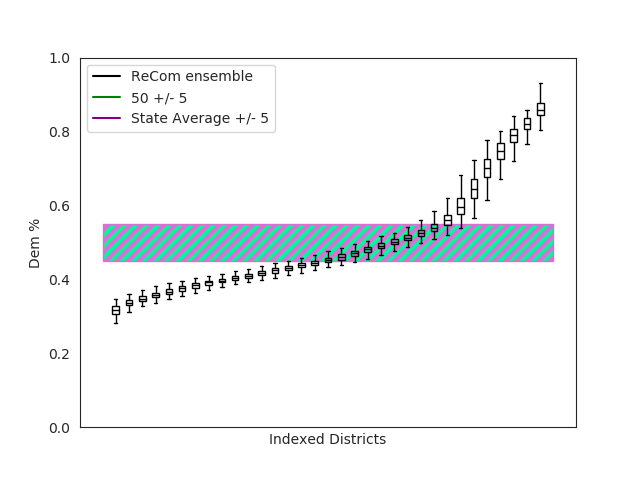}};    
\node at (5,4) {\includegraphics[width=160pt]{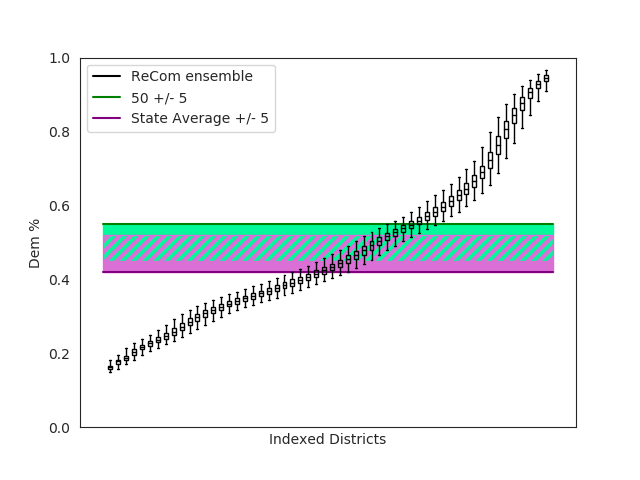}};    
\node at (5,2) {\includegraphics[width=160pt]{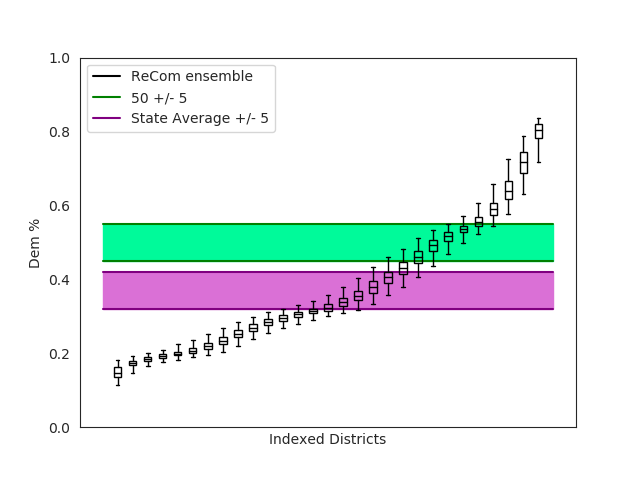}};    
\end{tikzpicture}         
\caption{Partisan share boxplots by district for the neutral ensembles in Congress (left) and state Senate (right).}
\label{fig:baselineBP}
\end{figure}

Even for a fixed distribution of votes, the district-by-district vote shares depend, of course, on the district lines.  
Figure \ref{fig:baselineBP}  shows the  distributions of the vote share across districts for each of the full ensembles.  In these plots, the districts in each plan are arranged from the one with the 
lowest Democratic vote share to the one with the highest.  The boxes show the 25th-75th percentile Democratic share, and the whiskers go from 1st-99th percentile.  Thus fatter boxes 
show a more elastic vote, with more control of partisan outcomes in the hands of the line-drawer.

In Table \ref{tab:avg50} we show the average number of $(5,50)$ and $(5, \Dbar)$ districts across the ensemble, compared to the enacted plans. The distribution of values observed across the ensemble  is similar to those of the enacted plans, with the exception of Georgia, where the ensemble finds many more districts in both bands.

\begin{table}[!h]
    \centering
    \begin{tabular}{|c||c|c||c|c|c|}
    \hline
 State, Type (\#)  &  Enacted & Ensemble &&Enacted& Ensemble \\
 {}&$(x,5,50)$& $(x,5,50)$& $\Dbar$& $(x,5, \Dbar)$& $(x,5, \Dbar)$\\
 \hline
 \hline
   MA Cong (9) & 0 &0.24 & 64.7 &5 &3.85 \\
   \hline
   VA Cong (11) & 3 &2.79 & 52.8 &2 &2.82 \\
   \hline   
   WI Cong (8) & 1 &1.76 & 49.6 &1 &1.81 \\
   \hline   
   GA Cong (14) & 2 &2.63 & 47.3 &2 &4.43 \\
   \hline
   UT Cong (4) & 1 &0.62 & 37.6 &2 &1.05 \\
   \hline   
   \hline
   MA Sen (40) & 6 &6.85 & 64.7&13 &12.82 \\
   \hline
   VA Sen (40) & 6 &8.60 &52.8 & 6 &8.00 \\
   \hline
   WI Sen (33) &7  &8.80& 49.6&7 &9.10 \\
   \hline
   GA Sen (56) & 2 &8.57 & 47.3& 4&8.78 \\
   \hline
   UT Sen (29) & 2 &4.09 & 37.6 &4 &4.57 \\
   \hline

    \end{tabular}
    \caption{Comparison of the number of $(5,50)$ and $(5, \Dbar)$ districts---close to 50-50 and close to the state average, respectively---in enacted plans versus ensemble means. Most cases, the enacted plan has slightly fewer $(5,50)$ districts than the ensemble
    average---the only examples where the enacted plan has more than the mean are Virginia's and Utah's Congressional plans.
Georgia and Utah's Senate numbers stand out from the group in the other direction, with conspicuously fewer close seats than typical plans in the neutral ensemble.}
    \label{tab:avg50}
\end{table}

The baselines reveal the extent to which the various states' political geographies shape the ensemble statistics. 
With the exception of Virginia, whose $MM$ distribution is centered at zero, all of the states have significantly Republican-favoring
average mean--median scores of between two and five percent.\footnote{The standard narrative around mean--median scores holds that $MM=.05$ means that the Republican party could expect half of the seats with only 45\% of the votes.}
For several ensembles, more than 90\% of the plans have a Republican-favoring score. This adds to a rapidly growing body 
of evidence that shows that applying uniform standards across states, without accounting for how voters are distributed throughout each state, can lead to misleading results and unattainable ideals. For example, a plan whose $MM$ value is typical for the  Wisconsin ensemble would be a significant outlier if evaluated against the Virginia data, even though the difference in statewide average vote share is small.\footnote{In a similar vein, there is growing evidence that the $MM$ score is not performing as advertised to signal meaningful partisan advantage \cite{PartSymm}.  Nonetheless,
we use it here to highlight that competitiveness may not be independent of other popular partisan
indicators.}

The seats outcomes let us measure the extent of partisan advantage that is entailed by political geography.
For instance, Wisconsin's vote pattern which was very close to 50-50 in Pres16,  but the  most Democratic-favoring plans found by the state Senate ensemble have only $14/33\approx 42\%$ Democratic seats. 

\subsection{Satisfiable vote-band rules}

\begin{figure}[!h]
    \centering
\begin{tikzpicture}[xscale=1.6,yscale=2.2]    
\node at (0,10) {\includegraphics[width=160pt]{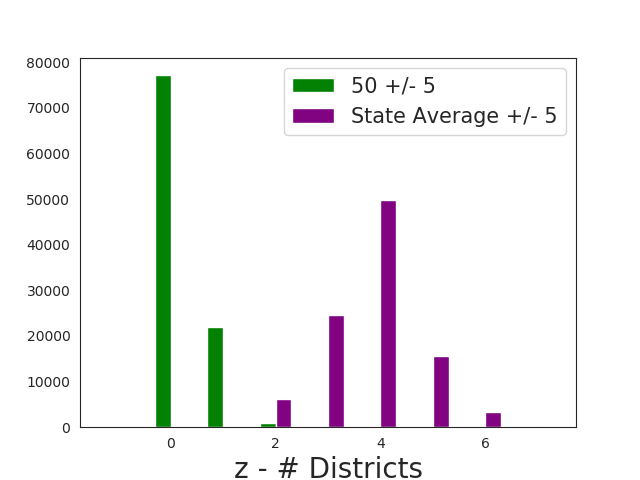}};    
\node at (0,8) {\includegraphics[width=160pt]{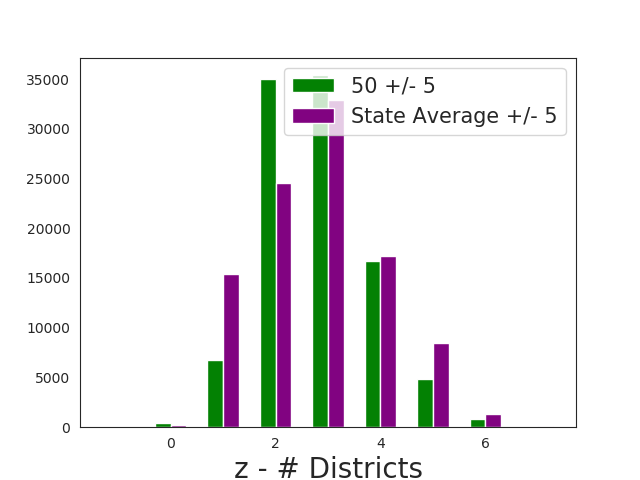}};    
\node at (0,6) {\includegraphics[width=160pt]{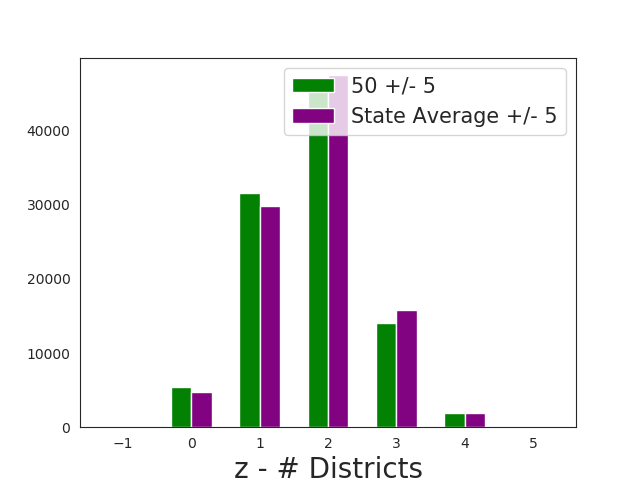}};    
\node at (0,4) {\includegraphics[width=160pt]{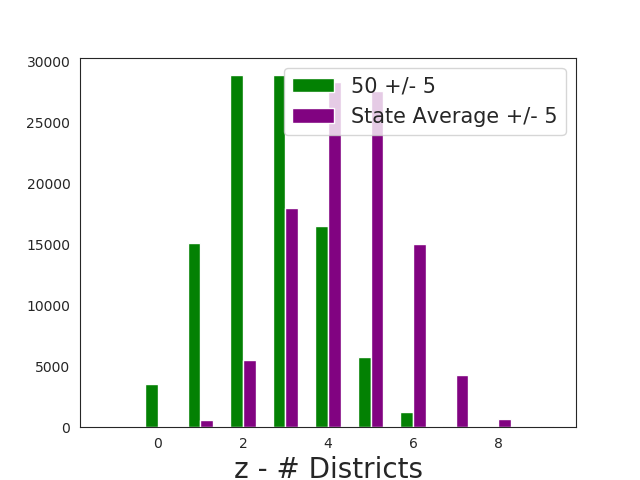}};    
\node at (0,2) {\includegraphics[width=160pt]{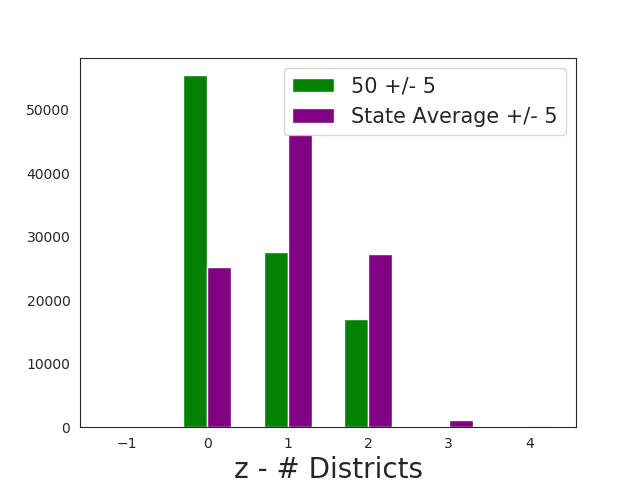}};  

\node at (2.5,10) {MA};
\node at (2.5,8) {VA};
\node at (2.5,6) {WI};
\node at (2.5,4) {GA};
\node at (2.5,2) {UT};

\node at (5,10) {\includegraphics[width=160pt]{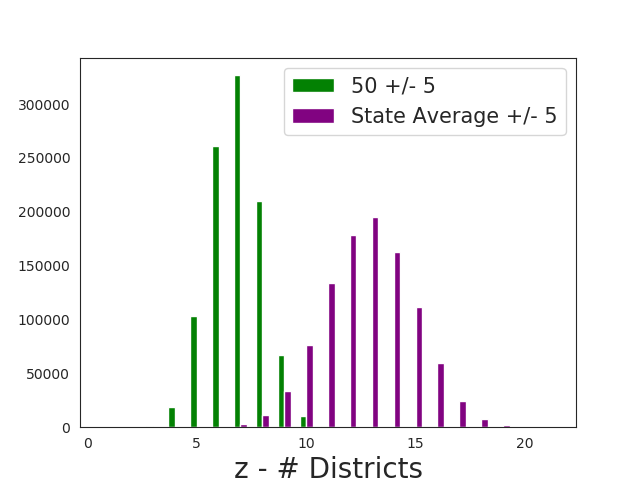}};    
\node at (5,8) {\includegraphics[width=160pt]{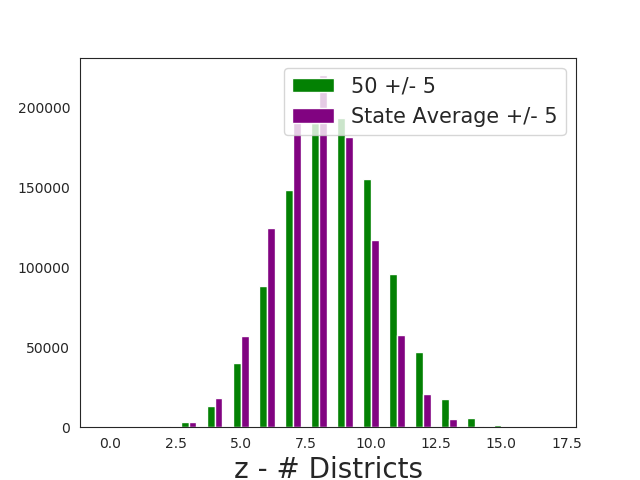}};    
\node at (5,6) {\includegraphics[width=160pt]{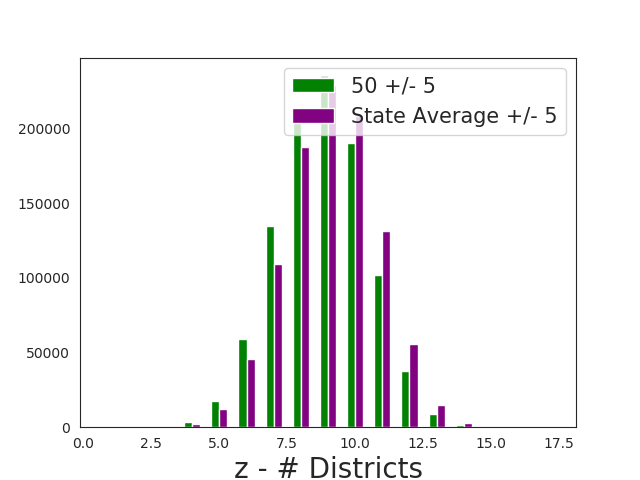}};    
\node at (5,4) {\includegraphics[width=160pt]{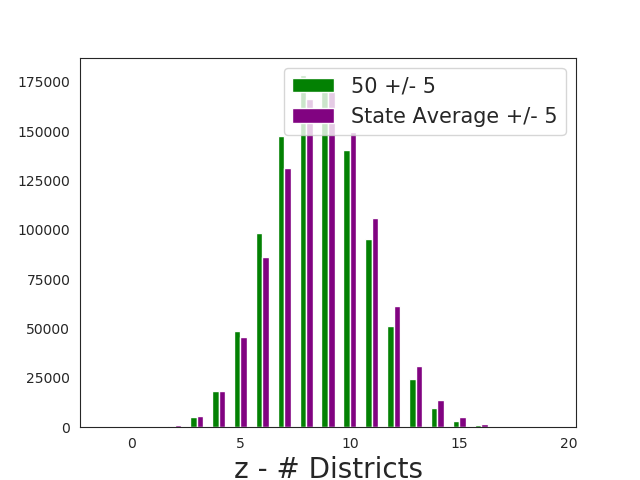}};    
\node at (5,2) {\includegraphics[width=160pt]{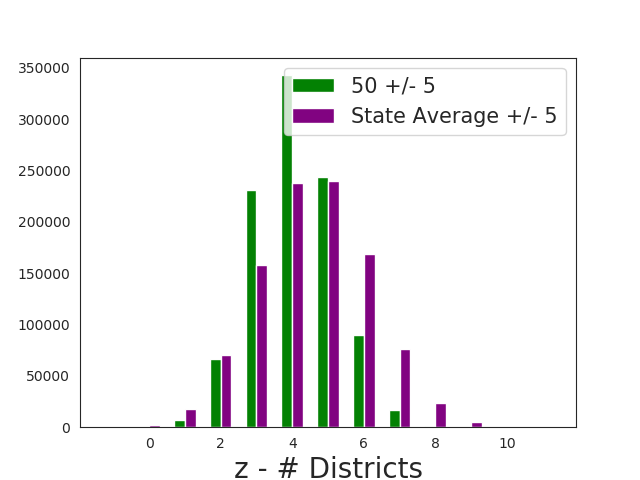}};    
\end{tikzpicture}       
\caption{How often each number of $(5,50)$ and $(5,\Dbar)$ districts was observed in the neutral ensemble.
Scale effects are notable in Georgia especially, where the competitiveness and state-typicality statistics
are the same for state Senate but noticeably different for Congress.}
\label{fig:basic-hists}
\end{figure}

\begin{figure}[!h]
    \centering
\begin{tikzpicture}[xscale=1.4,yscale=2.2]    
\node at (0,10) {\includegraphics[width=160pt]{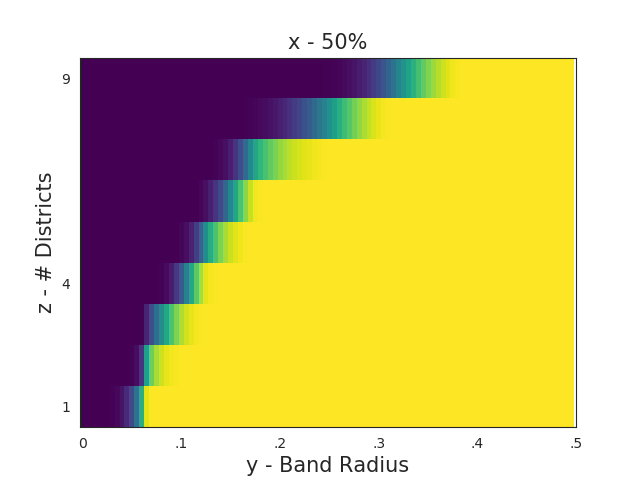}};
\node at (0,8) {\includegraphics[width=160pt]{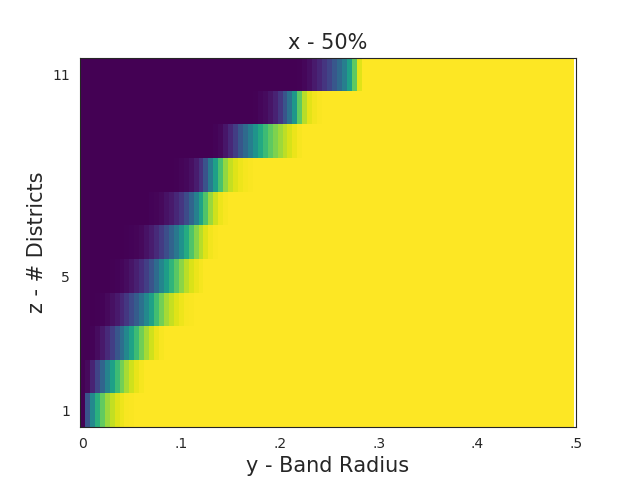}};    
\node at (0,6) {\includegraphics[width=160pt]{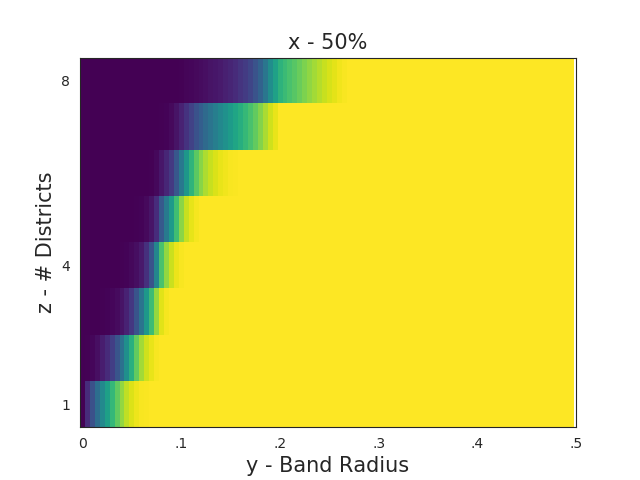}};    
\node at (0,4) {\includegraphics[width=160pt]{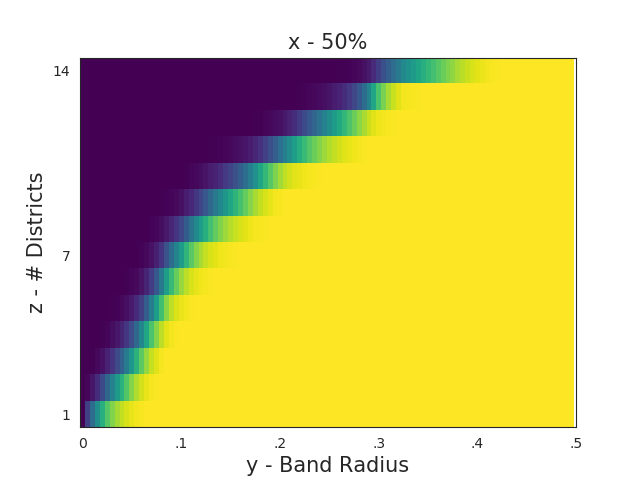}};    
\node at (0,2) {\includegraphics[width=160pt]{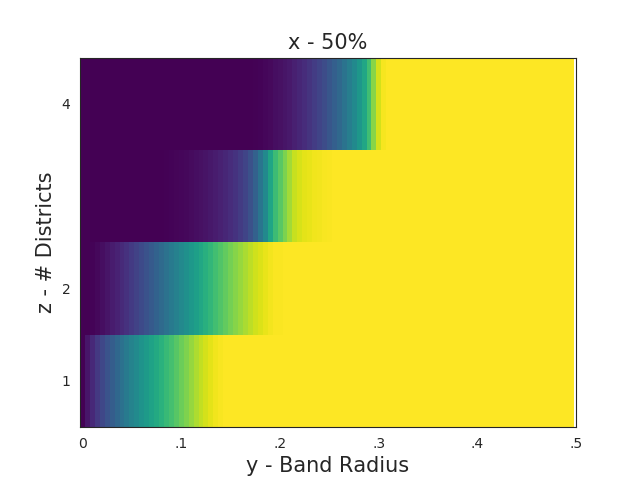}};  

\node at (2.5,10) {MA};
\node at (2.5,8) {VA};
\node at (2.5,6) {WI};
\node at (2.5,4) {GA};
\node at (2.5,2) {UT};

\node at (5,10) {\includegraphics[width=160pt]{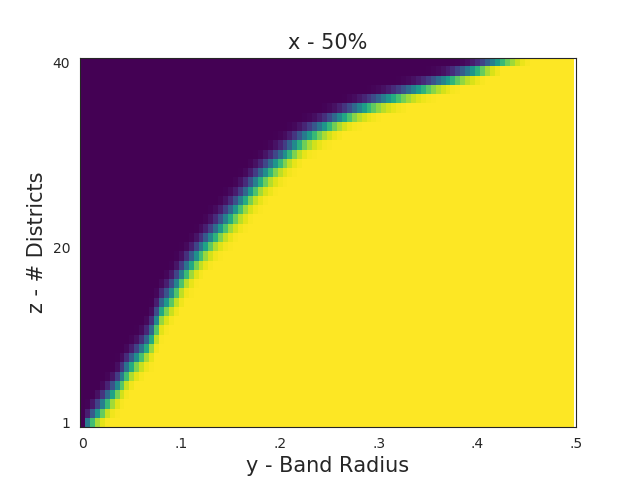}};    
\node at (5,8) {\includegraphics[width=160pt]{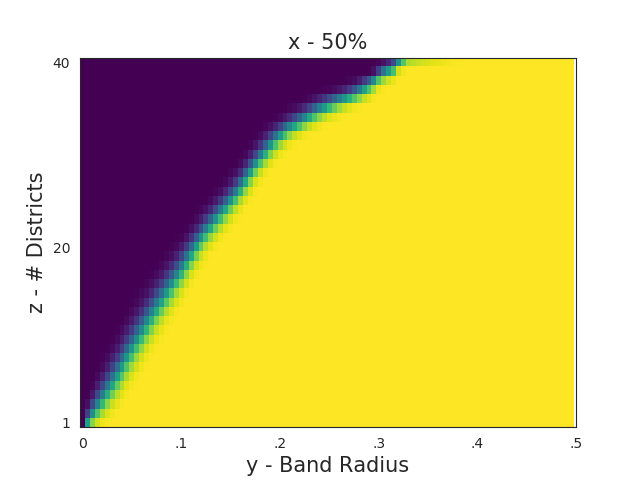}};    
\node at (5,6) {\includegraphics[width=160pt]{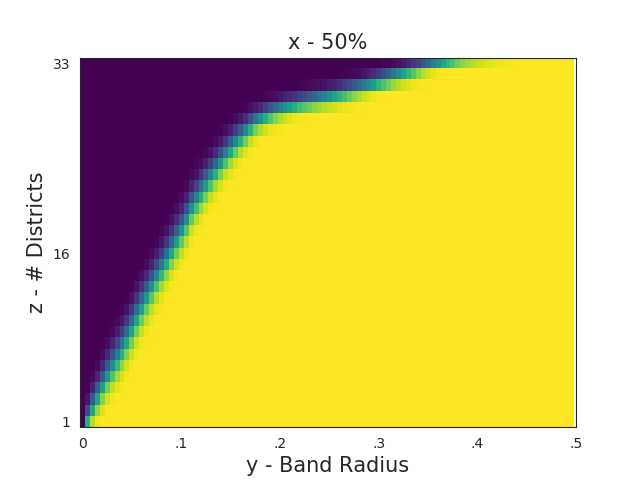}};    
\node at (5,4) {\includegraphics[width=160pt]{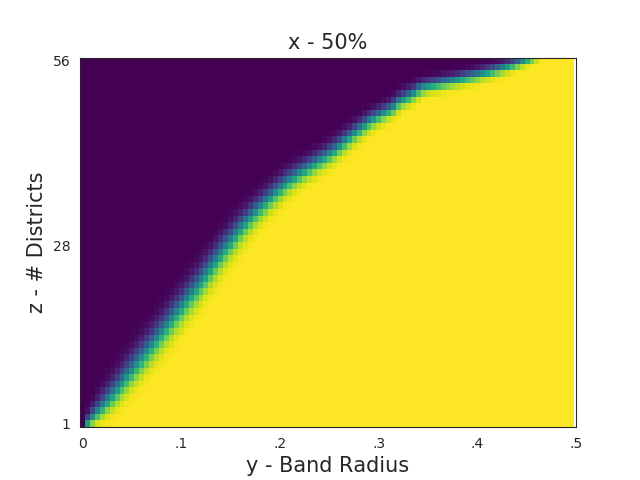}};    
\node at (5,2) {\includegraphics[width=160pt]{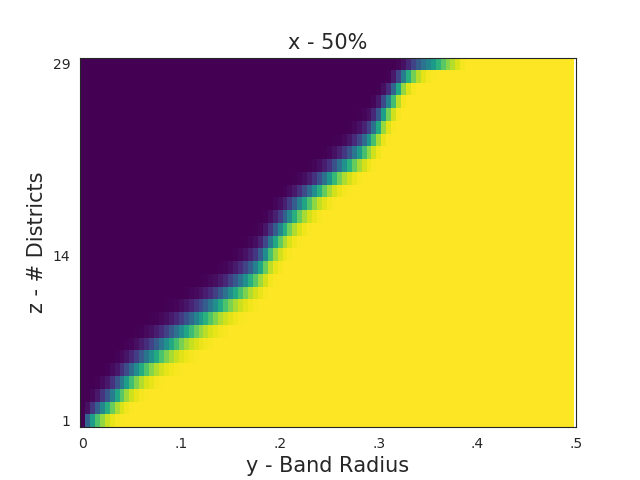}};    

\node at (9,6) {\includegraphics[width=100pt]{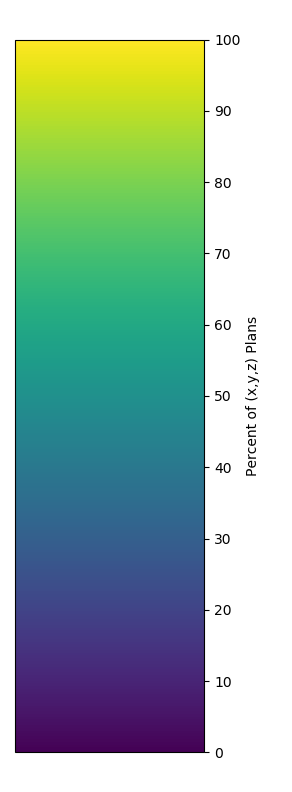}};
\end{tikzpicture}       
\caption{Visualizing the narrow range of vote-band rules that meaningfully constrain a neutral ensemble.  Congress (left), state Senate (right).}\label{fig:sometimes-zone}
\end{figure}

Too loose of a vote-band rule will impose no constraint at all on redistricting; too tight of a rule will be impossible to meet.  
In this section, we examine the level of constraint imposed by $(y,50)$ and $(y,\Dbar)$ rules.
This effect is explored in  Figures~\ref{fig:basic-hists}-\ref{fig:sometimes-zone}. 
Figure~\ref{fig:basic-hists} shows how often each number of competitive or state-typical districts occurs in a neutral ensemble.  This is intended to visualize the elasticity of the vote:
a skinny and tall histogram indicates that moving the lines does not allow much change to the count of a certain type of district, while a short and fat histogram indicates more freedom to change
the outcome by controlling the district lines.  
In Figure~\ref{fig:sometimes-zone}, the same data is presented in a new way to facilitate the interpretation of candidate rules.
The left column shows the existence and proportion of Congressional plans in the ensemble that are $(x,y,50)$ compressed, and the right  column shows similar plots for the state Senate. 
The transition regions for each fixed $x$ value are relatively small as a function of $y$;
but it is precisely these transition regions in which a vote-band rule would have any impact.  Outside these zones,
a vote-band requirement is either vacuous or is so hard to satisfy that a large ensemble finds no examples at all.  At a minimum, states must perform modeling of this kind using  recent electoral patterns before adopting a numerically specific vote-band rule for competitiveness or typicality.

For all of the states, the neutral ensemble found Congressional plans with no $(5,50)$ districts---that is, it is always possible to draw a plan with no districts at all between 45 and 55\% partisan share. 
Even within a fixed state, different vote elasticity may be observed at different scales.  
Comparing the behavior of Georgia's Congressional ensemble to the state Senate ensemble shows that requiring a fixed proportion of seats to be competitive would impose a much more constraining condition on the Congressional districts than the state Senate.

\subsection{Partisan impacts of vote-band winnowing}
\label{sec:winnow}

One way that vote-band rules might play out is that among contending plans, those that best satisfy the rules would be selected. 
To model this scenario, we now extract subsets of plans that are $(x,y,z)$ compressed and compare the aggregate properties of these plans to the full ensemble distributions.  
Comparing the subset of an ensemble that has a certain property with the full set from which it was drawn is known as {\em winnowing}.
This allows us to observe whether the competitive plans differ from the neutral ensemble in a systematic fashion under the partisan metrics. 

\begin{figure}[!h]
    \centering
\begin{tikzpicture}[xscale=1.8,yscale=1.8]    
\node at (0,10) {\includegraphics[width=140pt]{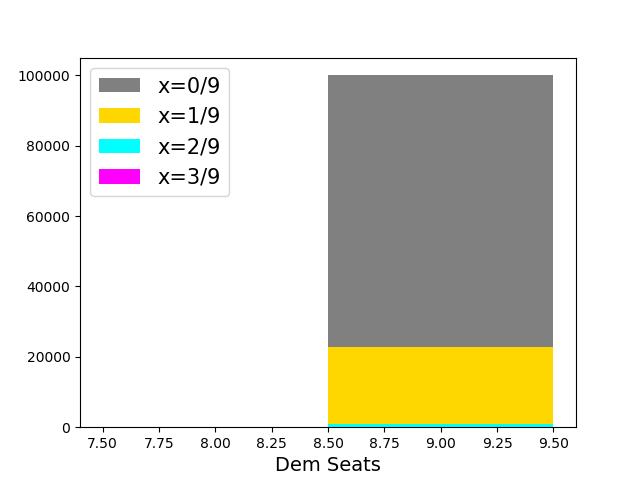}};    
\node at (0,8) {\includegraphics[width=140pt]{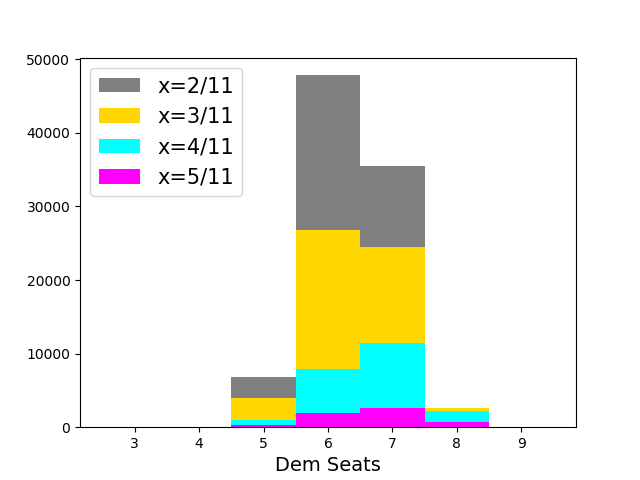}};    
\node at (0,6) {\includegraphics[width=140pt]{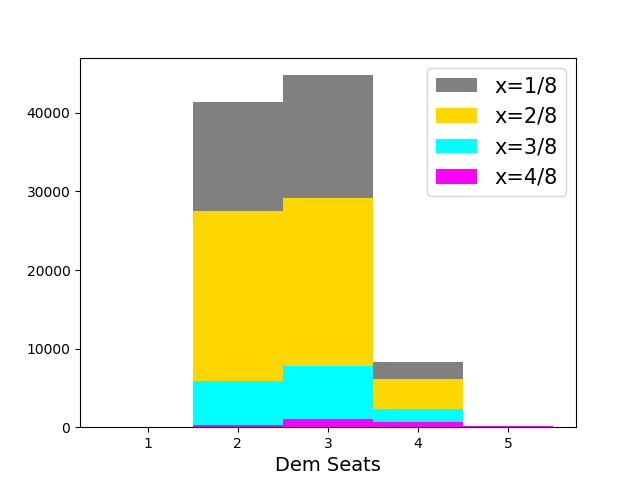}};    
\node at (0,4) {\includegraphics[width=140pt]{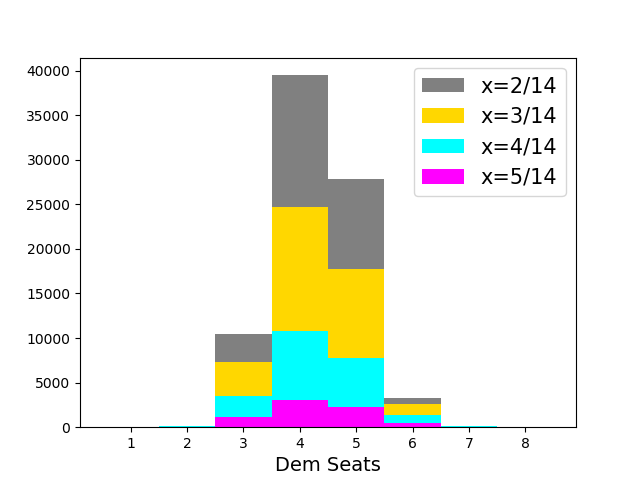}};    
\node at (0,2) {\includegraphics[width=140pt]{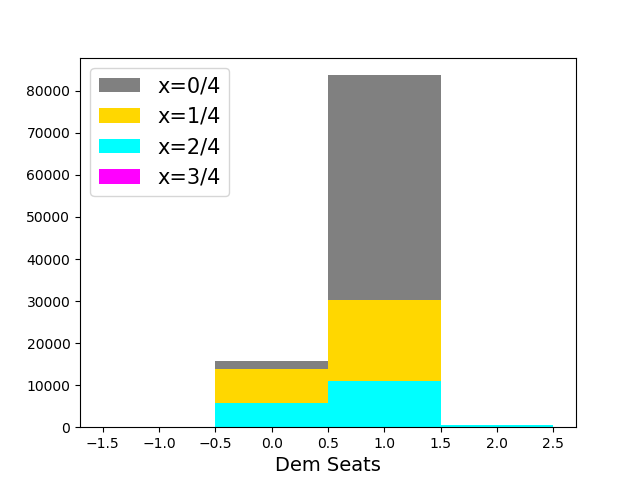}};  

\node at (2.5,10) {\includegraphics[width=140pt]{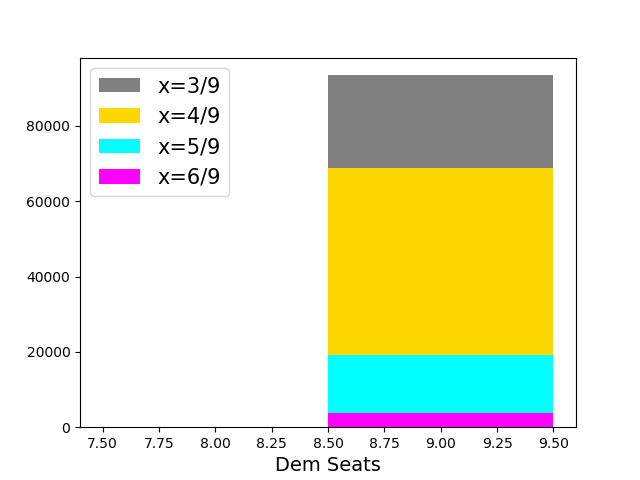}};    
\node at (2.5,8) {\includegraphics[width=140pt]{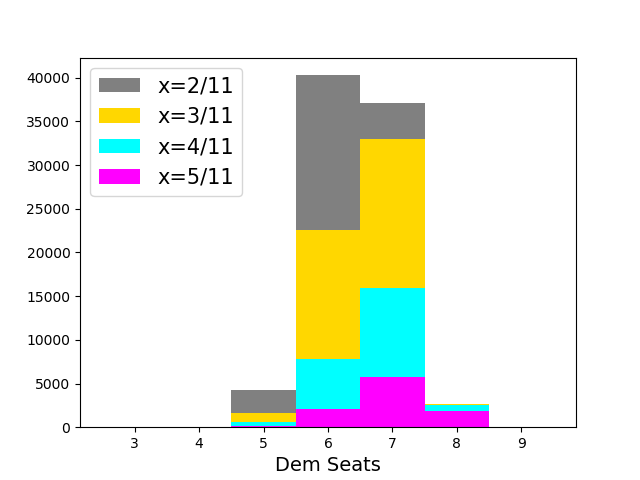}};    
\node at (2.5,6) {\includegraphics[width=140pt]{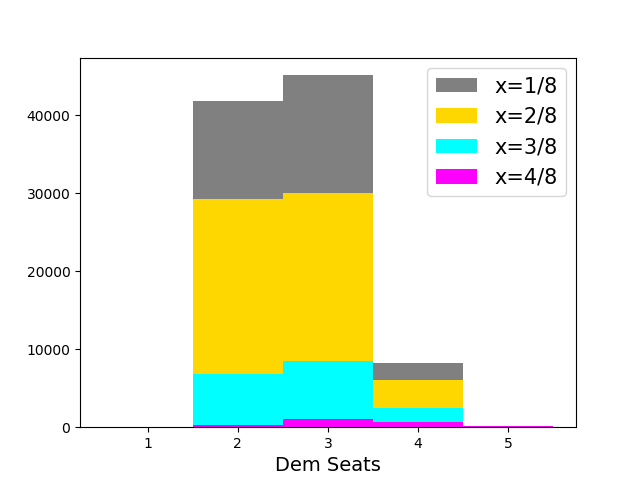}};    
\node at (2.5,4) {\includegraphics[width=140pt]{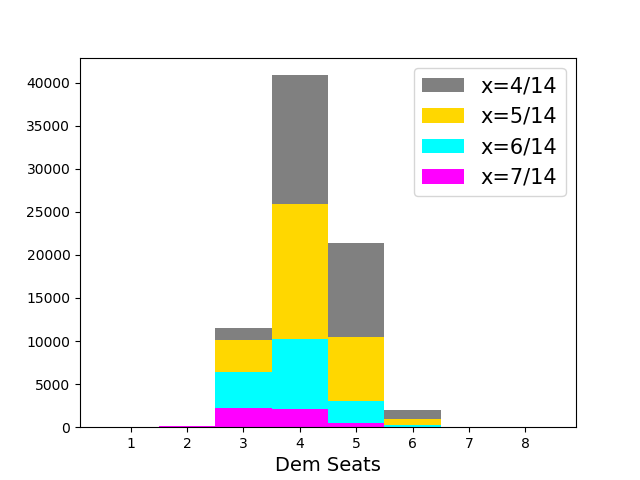}};    
\node at (2.5,2) {\includegraphics[width=140pt]{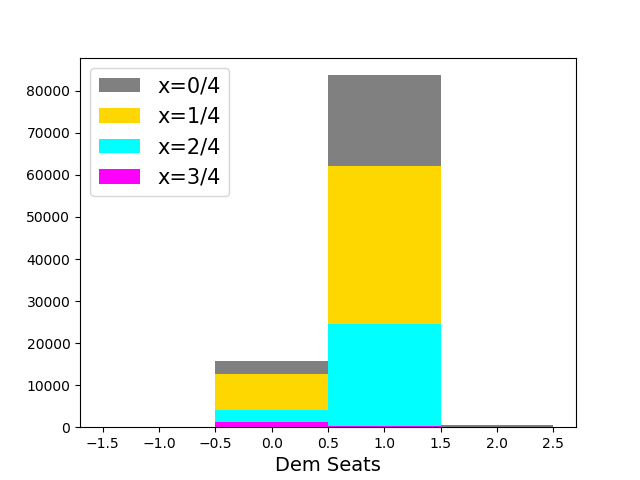}};  

\node at (5,10) {\includegraphics[width=140pt]{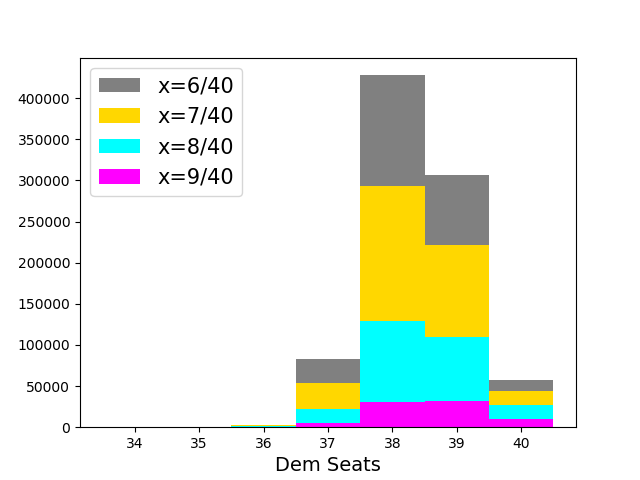}};    
\node at (5,8) {\includegraphics[width=140pt]{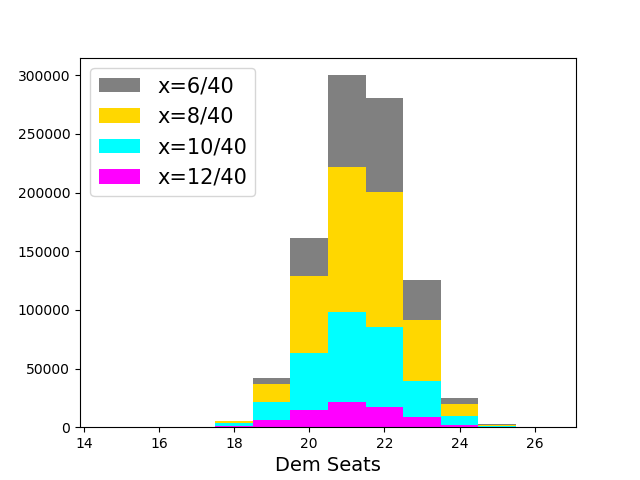}};    
\node at (5,6) {\includegraphics[width=140pt]{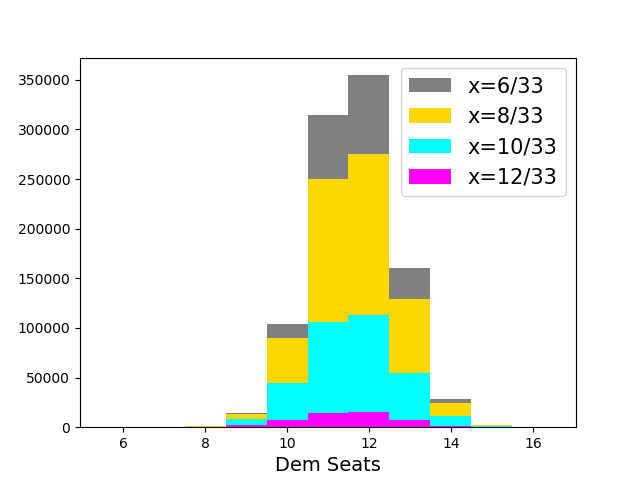}};    
\node at (5,4) {\includegraphics[width=140pt]{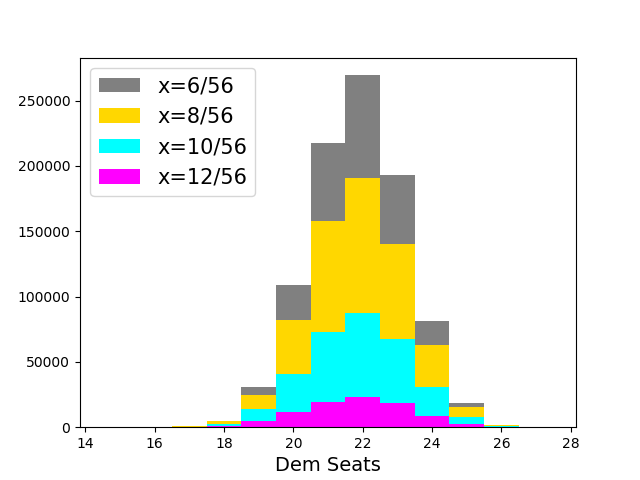}};    
\node at (5,2) {\includegraphics[width=140pt]{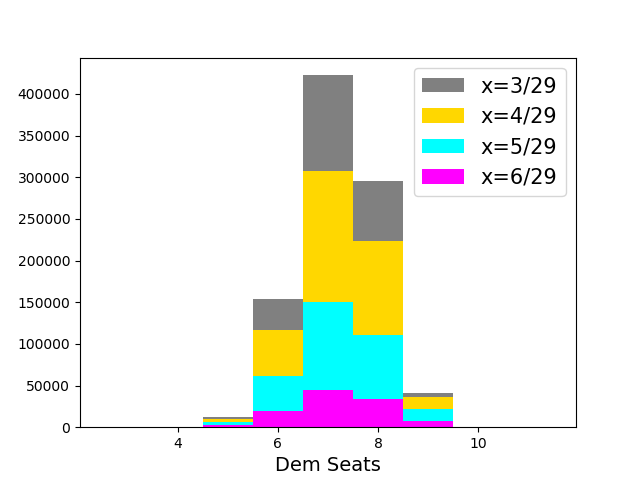}};    

\node at (7.5,10) {\includegraphics[width=140pt]{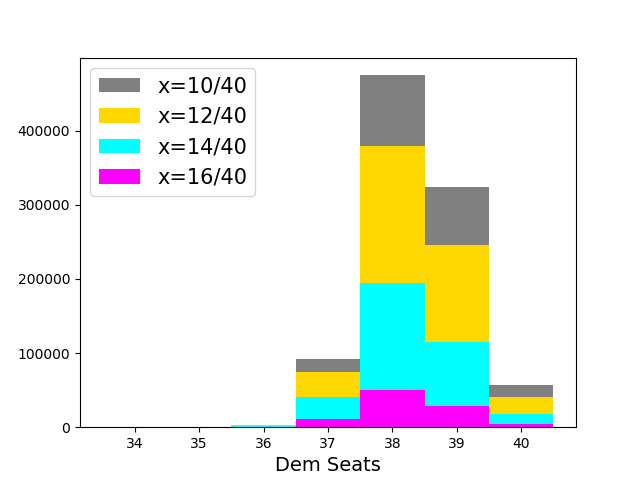}};    
\node at (7.5,8) {\includegraphics[width=140pt]{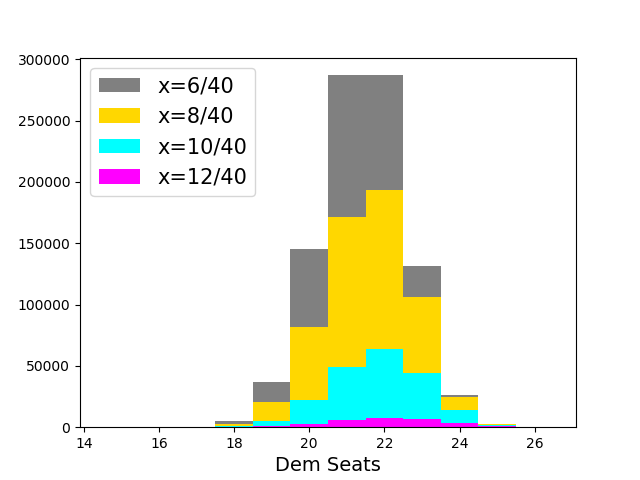}};    
\node at (7.5,6) {\includegraphics[width=140pt]{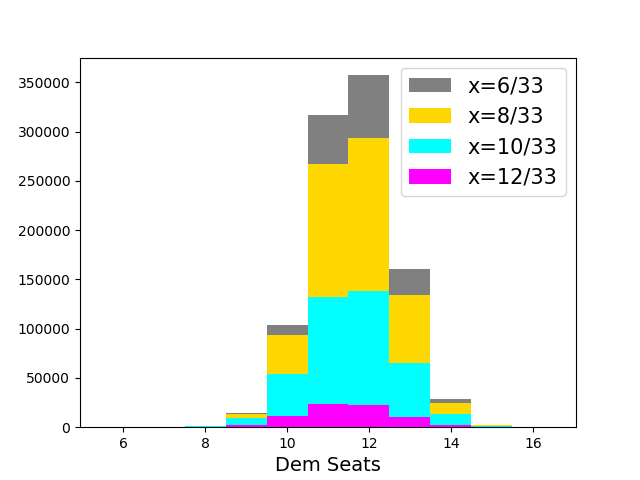}};    
\node at (7.5,4) {\includegraphics[width=140pt]{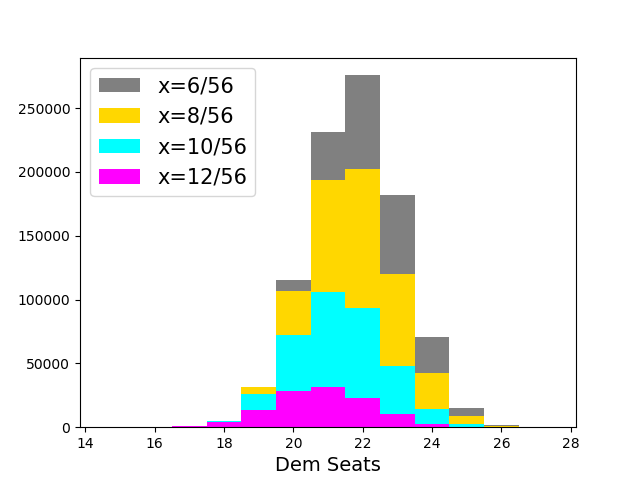}};    
\node at (7.5,2) {\includegraphics[width=140pt]{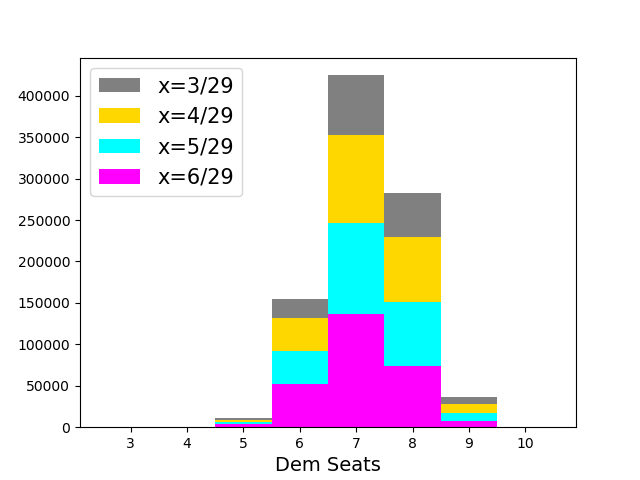}};  
\end{tikzpicture}          
\caption{Histograms of  Democratic seat outcomes in  $(x,y,z)$ compressed plans as a function of increasing $x$. Columns 1-2: Congress, $z=50$ and $z=\Dbar$.  Columns 3-4: state Senate,
$z=50$ and $z=\Dbar$. States, top to bottom: 
MA, VA, WI, GA, UT.}
\label{fig:onlydemseats}
\end{figure}

We begin by computing the average number of Democratic seats in districting plans that are $(x,5,50)$ and $(x,5,\Dbar)$ compressed, as a function of $x$. Figure \ref{fig:onlydemseats} shows these results for each state we analyzed.  For states with a strong statewide party skew, increasing the number of  districts close to the statewide average unsurprisingly tends to shift the seats outcome in favor of the party with the larger statewide percentage.  In tilted states, however, vying for competitiveness can actually 
hurt the minority party.  
The Massachusetts state Senate results show an example of a potentially surprising consequence of enforcing a competitiveness constraint. For this ensemble, increasing the number of $(5,50)$   districts increases the expected number of Democratic seats. This is because it is easier to create more seats with a Democratic share of roughly 55\% by taking scarce Republican votes from the few Republican-leaning districts.  

The case of Wisconsin ($\Dbar\approx 50$) is particularly interesting, 
since a vote-band competitive rule for Congress imposes a Democratic advantage, while a similar rule for state Senate 
has a Republican lean.

\begin{figure}[!h]
    \centering
\begin{tikzpicture}[xscale=1.8,yscale=1.8]    
\node at (0,10) {\includegraphics[width=130pt]{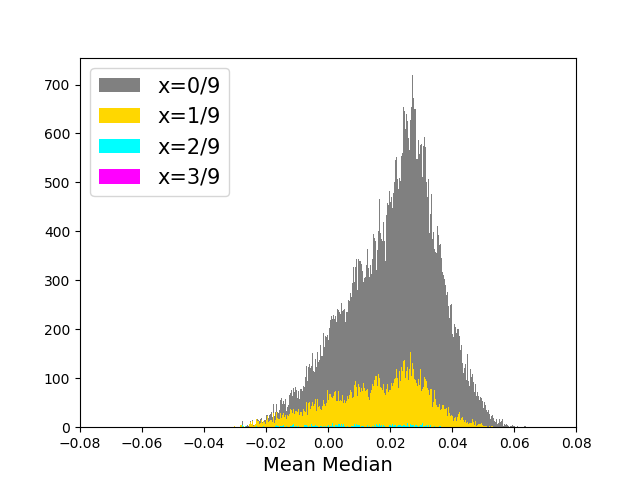}};    
\node at (0,8) {\includegraphics[width=130pt]{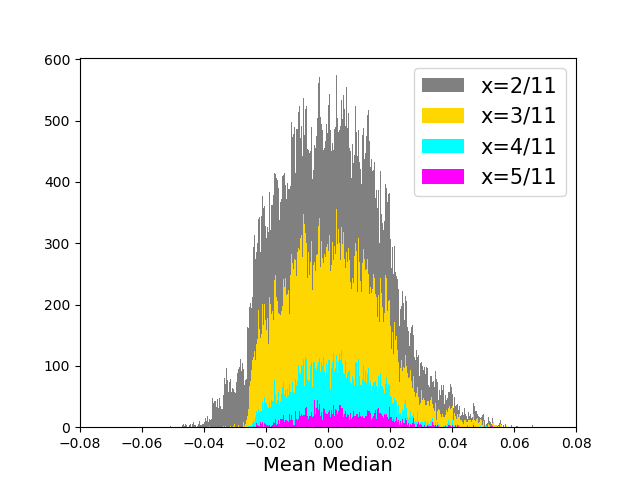}};    
\node at (0,6) {\includegraphics[width=130pt]{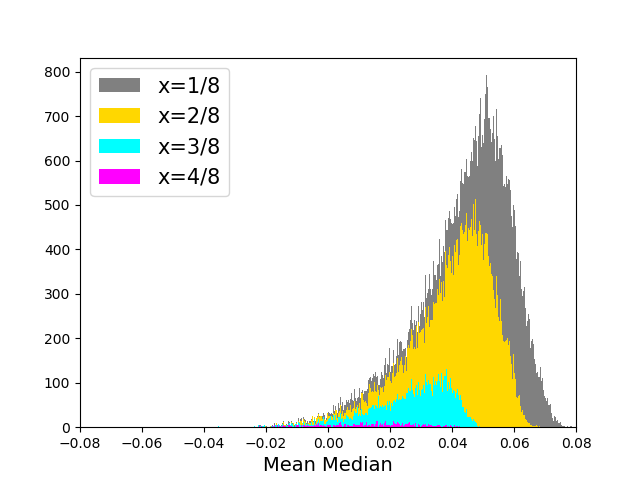}};    
\node at (0,4) {\includegraphics[width=130pt]{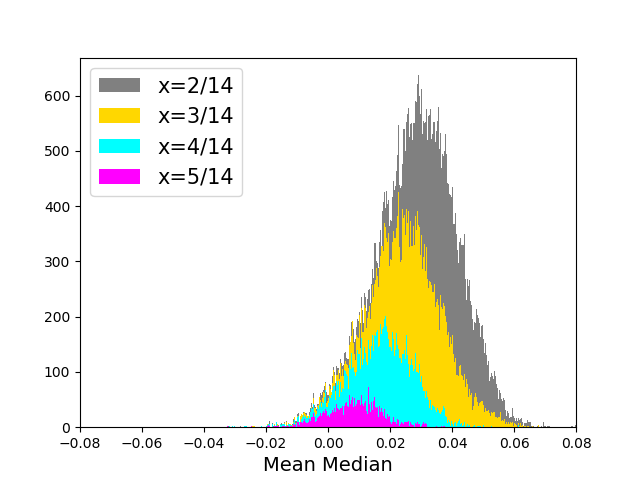}};    
\node at (0,2) {\includegraphics[width=130pt]{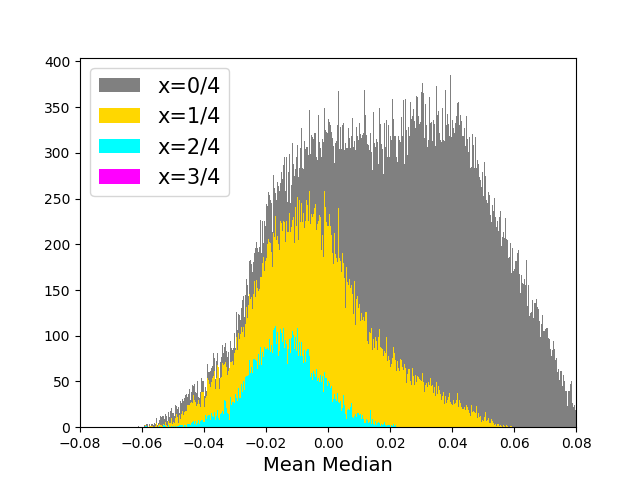}};  

\node at (2.5,10) {\includegraphics[width=130pt]{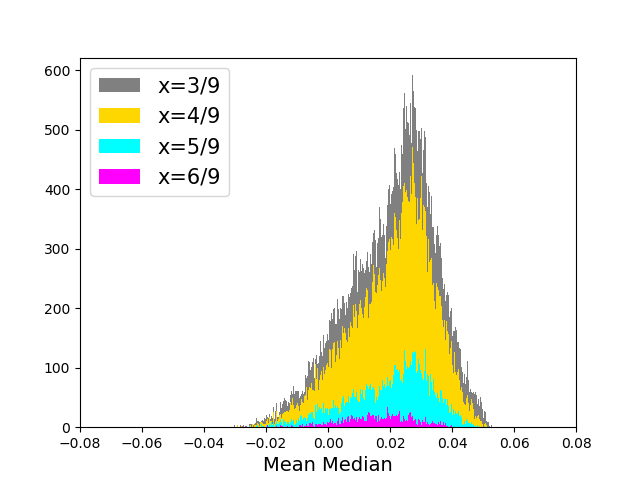}};    
\node at (2.5,8) {\includegraphics[width=130pt]{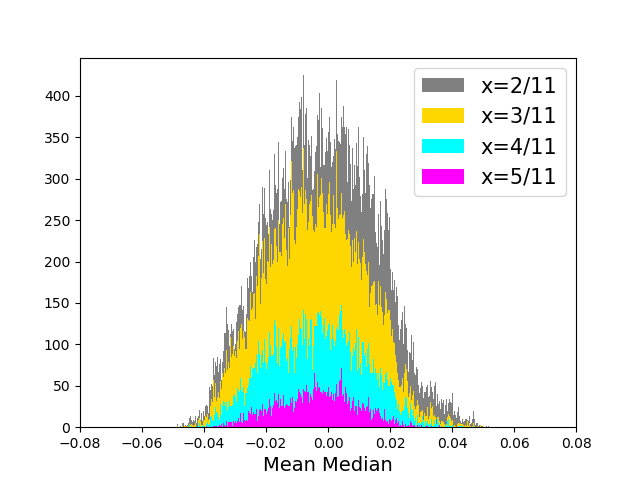}};    
\node at (2.5,6) {\includegraphics[width=130pt]{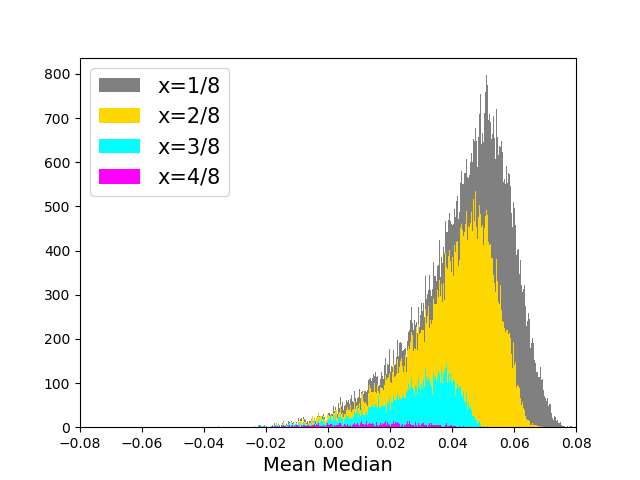}};    
\node at (2.5,4) {\includegraphics[width=130pt]{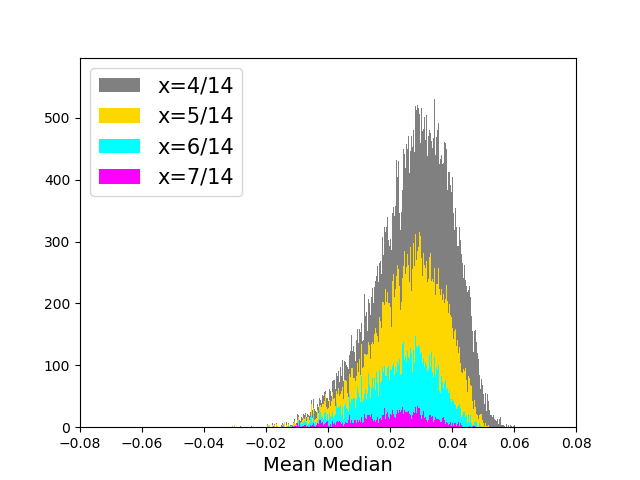}};    
\node at (2.5,2) {\includegraphics[width=130pt]{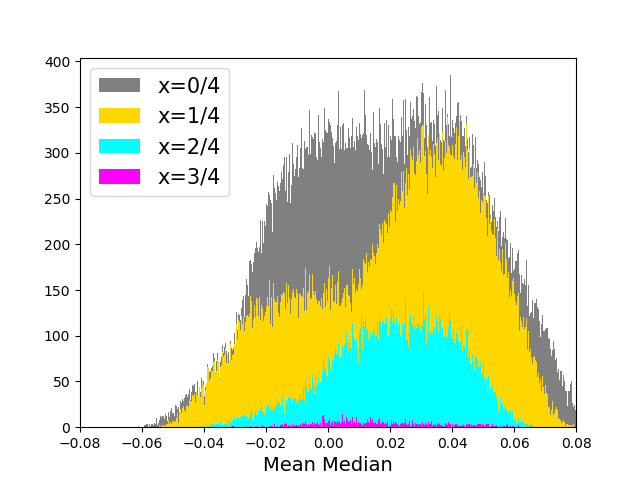}};  

\node at (5,10) {\includegraphics[width=130pt]{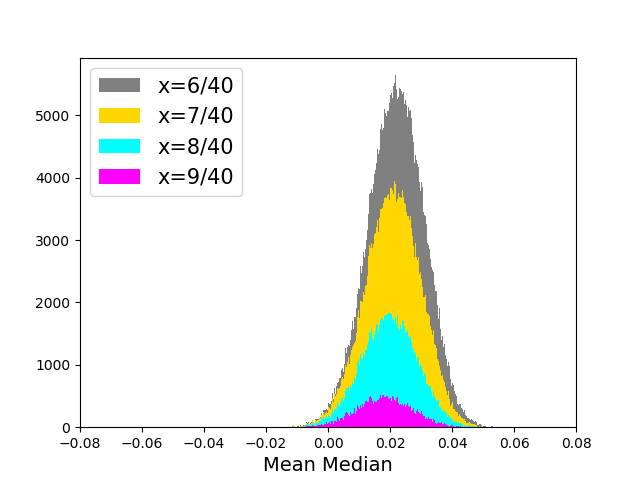}};    
\node at (5,8) {\includegraphics[width=130pt]{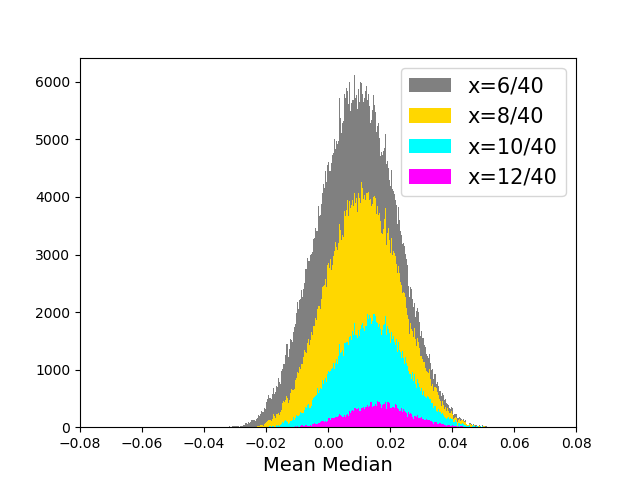}};    
\node at (5,6) {\includegraphics[width=130pt]{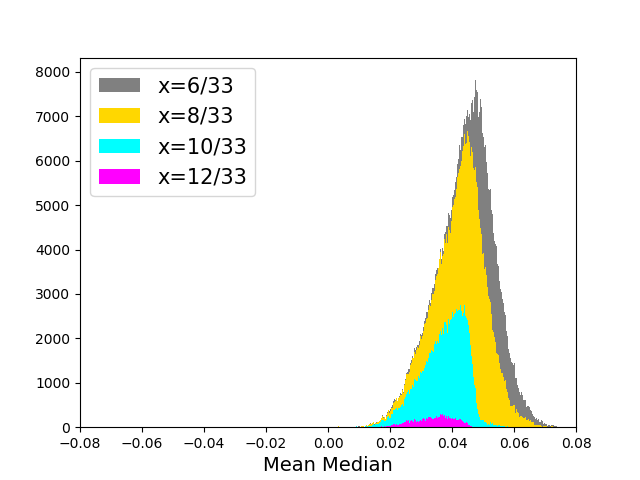}};    
\node at (5,4) {\includegraphics[width=130pt]{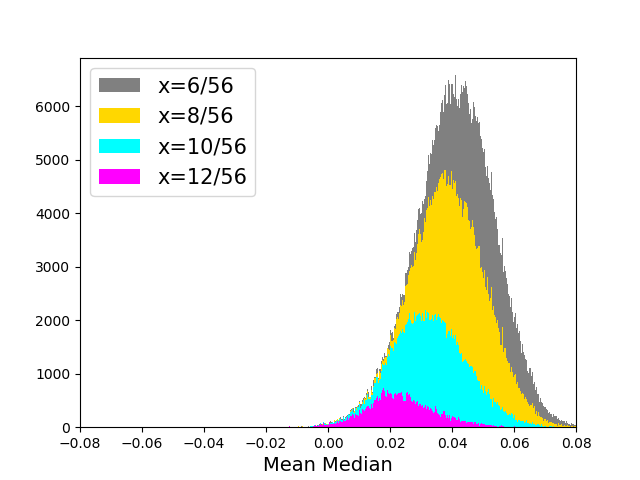}};    
\node at (5,2) {\includegraphics[width=130pt]{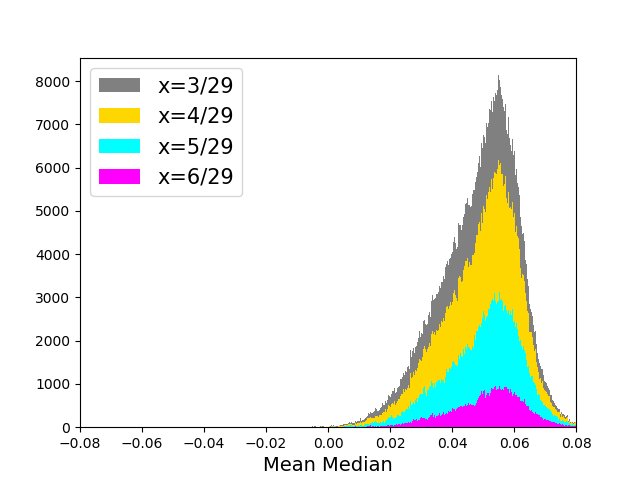}};    

\node at (7.5,10) {\includegraphics[width=130pt]{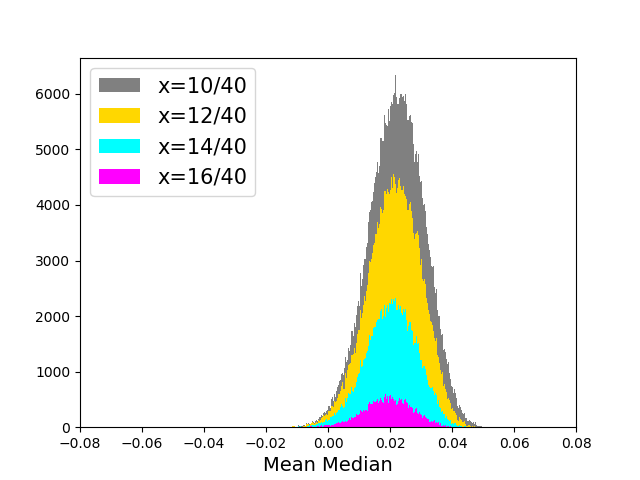}};    
\node at (7.5,8) {\includegraphics[width=130pt]{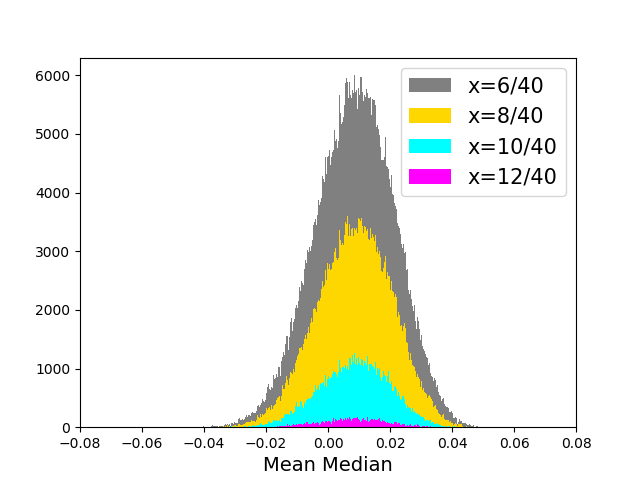}};    
\node at (7.5,6) {\includegraphics[width=130pt]{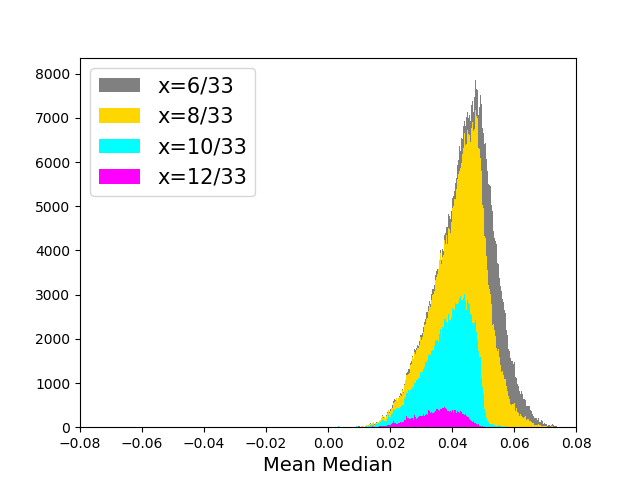}};    
\node at (7.5,4) {\includegraphics[width=130pt]{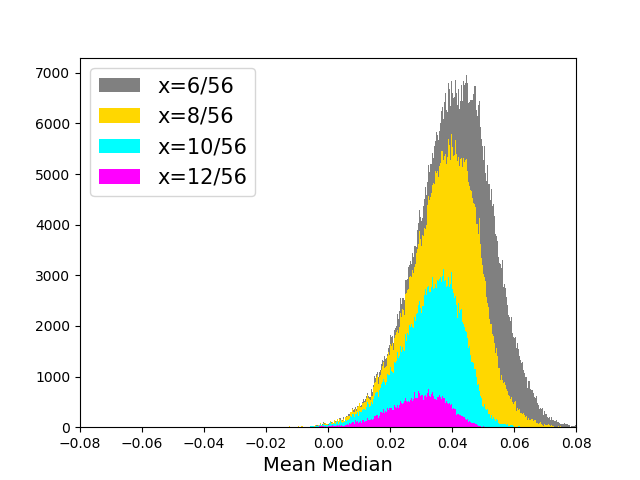}};    
\node at (7.5,2) {\includegraphics[width=130pt]{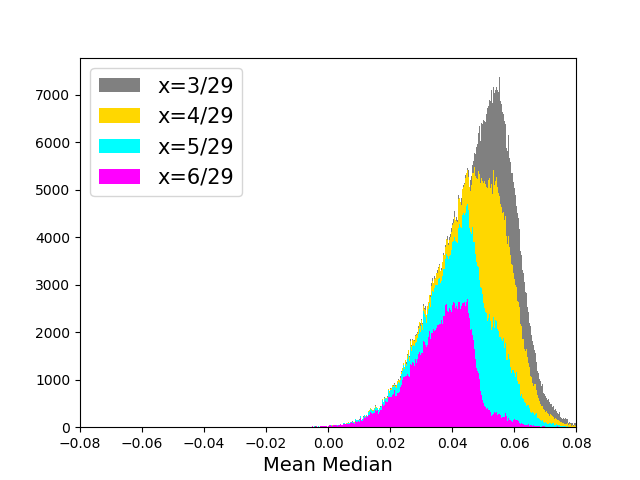}};  
\end{tikzpicture}             
\caption{Successively winnowed mean-median distributions for $(x,5,z)$ compressed plans as a function of increasing $x$. This figure shows how the MM score reacts as more districts are required to be competitive or state-typical.  (The grey histogram is the full ensemble.)
Columns 1-2: Congress, $z=50$ and $z=\Dbar$.  Columns 3-4: state Senate,
$z=50$ and $z=\Dbar$. States, top to bottom: 
MA, VA, WI, GA, UT.    
Recall that $MM=0$ is thought to signal symmetry between the parties, and positive values
are thought to signify Republican advantage.
One thing to note is that  all states except Virginia show a baseline of pronounced Republican advantage by this measure in the neutral ensemble (i.e., with no partisan gerrymandering).}
\label{fig:mm50state}
\end{figure}

Next, we compare the distributions of mean-median values on the winnowed sub-ensembles, shown in Figure \ref{fig:mm50state}. Again, the histograms show a  range of different behaviors as the constraints are tightened; several cases are worth highlighting. For the case $z=50$, both Wisconsin and Georgia, whose full ensemble averages are Republican-favoring, are forced towards the center as the constraint tightens, while Virginia, which was balanced in the full ensemble, becomes more Republican-favoring in both the Congressional and state Senate ensembles under progressively tighter winnowing. Contrast this behavior with Utah, where the winnowed Congressional ensemble leans approximately four points towards Democratic-favoring scores, while the state Senate ensemble looks even more extremely Republican-favoring.

\subsection{Partisan impacts of heuristic optimization}
\label{sec:opt}
While the previous trials investigated  the properties of competitive plans that were observed in a neutral ensemble, laws mandating competitiveness will also incentivize mapmakers to \emph{optimize} for competitive plans. Plans constructed with this aim in mind likely will be significantly different than those drawn without access to partisan information. To expose some of these differences, we use the partisan-aware optimization methods {\sf Opt1} and {\sf Opt2} (described in \S\ref{sec:methods}) to construct plans that have significantly  more $(5,50)$ competitive districts than were found by the neutral ensemble. 

To give an extremely crude sense of the success of these short runs with simple optimization heuristics,  we compare our  optimization runs to those expressly constructed for the 538 Atlas of Redistricting to minimize CPVI values (Table \ref{tab:538}). The comparison is not perfect, since the underlying data, models, and evaluation mechanism differ.\footnote{For example, the CVPI values use a slightly Democratic-favoring baseline as a result of the popular vote in the 2016 presidential election, which makes it slightly easier to make competitive districts in Democratic-leaning states and slightly harder in Republican-leaning ones. Additionally, the 538 maps had constraints on the number of majority-minority districts.}
Also, there are states in which one of the styles of optimization is a bad fit for the task at hand, like {\sf Opt2} in Massachusetts.
The similarity between our results and theirs, however, suggests that our simple optimization heuristics have some power.
An interesting direction for future work would be to use the precinct-sorting methods from \cite{MGGGMA} to get rigorous upper bounds on the possible number of competitive districts.

\begin{table}[!h]
    \centering
    {\footnotesize
    \begin{tabular}{|c||c||c||c|c||c|c|}
    \hline
 State, Type, &  538 Atlas of&Neutral& {\sf Opt1} & {\sf Opt1}& {\sf Opt2} & {\sf Opt2} \\
 (Total \#)& Redistricting& Max& Max&  Mean& Max &  Mean\\
 \hline
 \hline
   MA Cong (9) & {\bf 4} & 2&3& 2.11&1&0.05  \\
   \hline
     VA Cong (11) &{\bf 9} &7 & 8 &7.33&{\bf 9}&6.61\\
   \hline
   WI Cong (8) & 6 &5&6 &3.99&{\bf 8}&4.28 \\
   \hline   
    GA Cong (14)  & 7 &8&12 & 9.41&{\bf 14}&10.68 \\
   \hline  
   UT Cong (4) & 1&{\bf2} &{\bf2}  & 1.99&{\bf 2}&.51\\
   \hline
   \hline
   MA Sen (40) & -- &12 & {\bf 17}&15.35&9&5.12 \\
   \hline
   VA Sen (40)& -- &17  &26 &23.63&{\bf 28}&20.32 \\
   \hline
   WI Sen (33) &-- &17 &20 &17.33&{\bf 25}&15.83 \\
   \hline
   GA Sen (56)&--&19  &{\bf37} &32.99&31&20.85 \\
   \hline
   UT Sen (29) &-- &9 & {\bf 14}  & 11.42&13&6.93\\
   \hline
    \end{tabular}
    }
    \caption{Comparison of the number of $(5,50)$ districts found in the neutral and two styles of hill-climbing ensembles to the 538 Atlas of Redistricting's highly competitive plans. Maximum values highlighted in bold. In most cases, hill-climbing found plans where at least half the districts are $(5,50)$ competitive, which differs significantly from the results of the neutral ensemble.}
    \label{tab:538}
\end{table}

Comparing to the box plots of Figure \ref{fig:baselineBP}  gives a measure of how far these optimized plans are from those in the neutral ensemble. For Georgia Congressional plans, the 1\%-99\% whiskers of the box plots only hit the $(5,50)$ band for five out of 14 districts, but the optimization method found plans where all 14 districts are in the $(5,50)$ band. 
Once again, we find Georgia to be particularly elastic in its vote distribution and to have enacted relatively uncompetitive plans. Note we have
not made an effort to include the full suite of Georgia districting criteria in our model, and the Voting Rights Act in particular might be quite constraining.

\begin{figure}[!h]
    \centering
\begin{tikzpicture}[xscale=1.6,yscale=2.2]    
\node at (0,10) {\includegraphics[width=160pt]{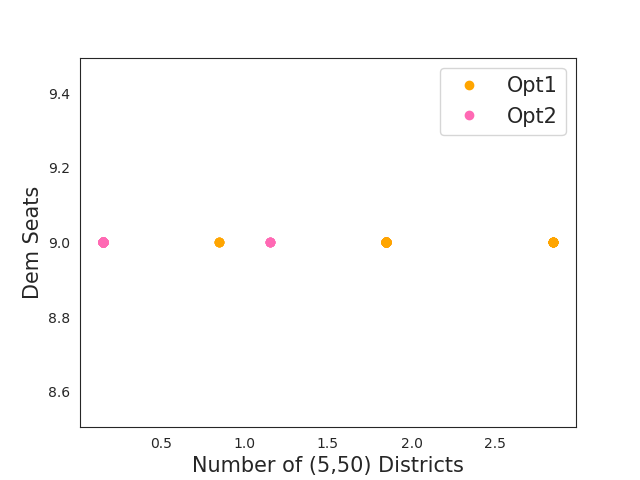}};    
\node at (0,8) {\includegraphics[width=160pt]{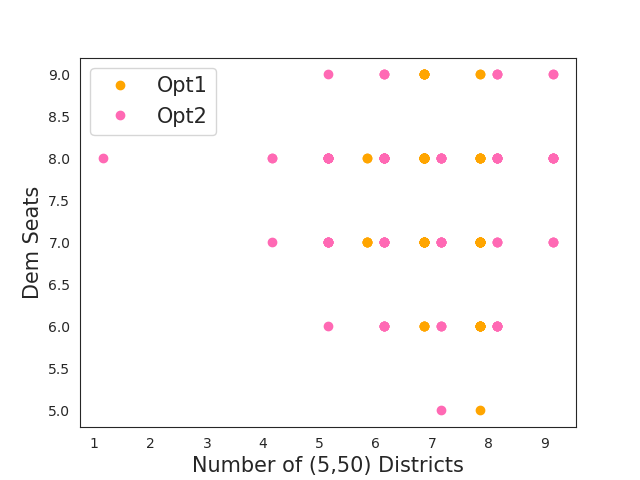}};    
\node at (0,6) {\includegraphics[width=160pt]{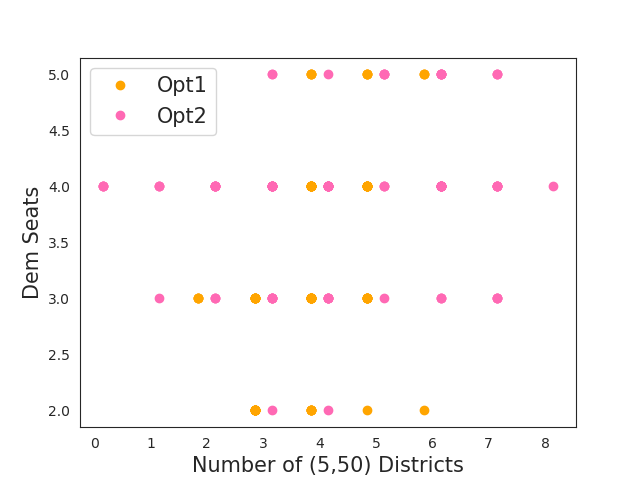}};    
\node at (0,4) {\includegraphics[width=160pt]{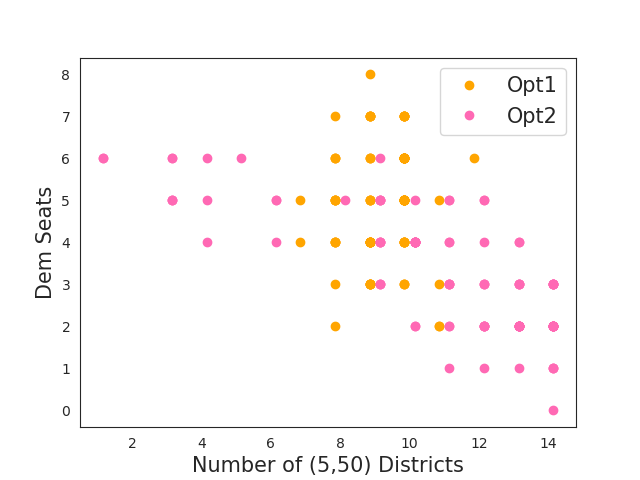}};    
\node at (0,2) {\includegraphics[width=160pt]{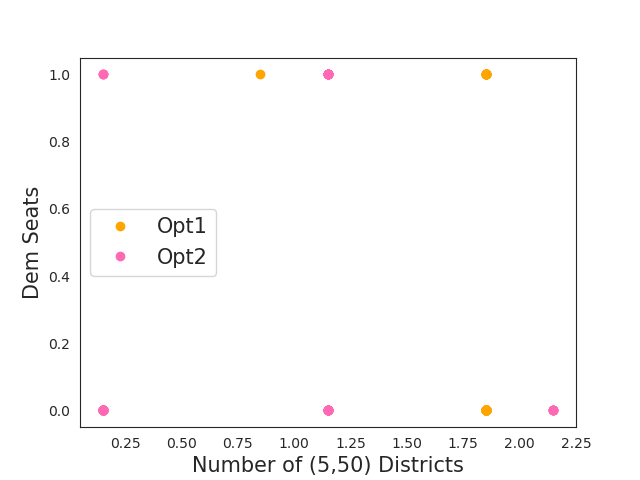}};  

\node at (2.5,10) {MA};
\node at (2.5,8) {VA};
\node at (2.5,6) {WI};
\node at (2.5,4) {GA};
\node at (2.5,2) {UT};

\node at (5,10) {\includegraphics[width=160pt]{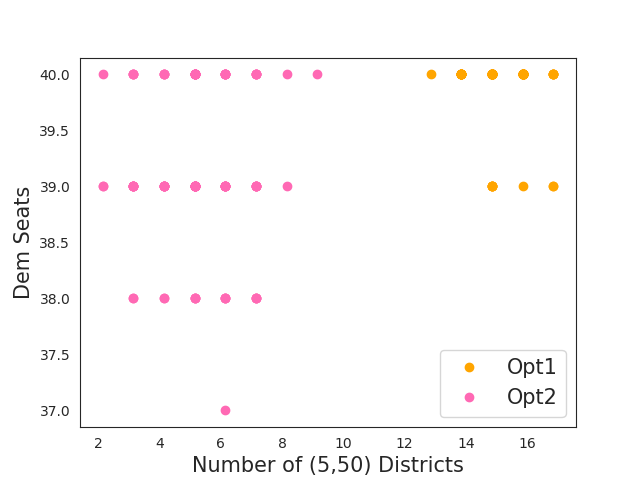}};    
\node at (5,8) {\includegraphics[width=160pt]{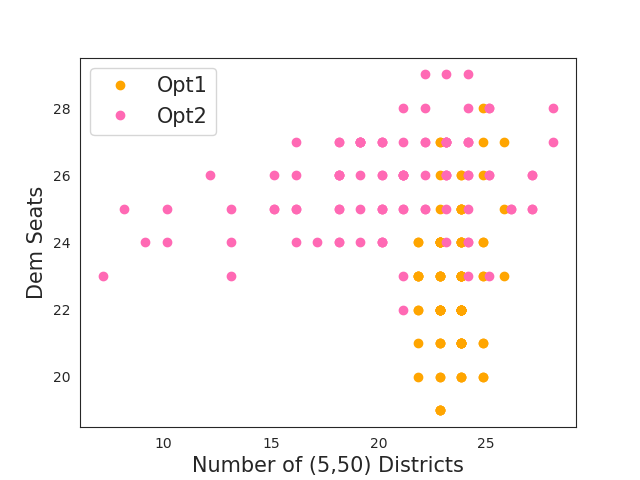}};    
\node at (5,6) {\includegraphics[width=160pt]{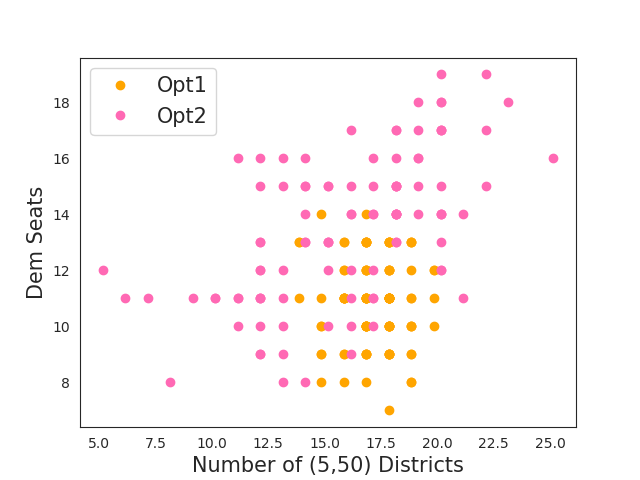}};    
\node at (5,4) {\includegraphics[width=160pt]{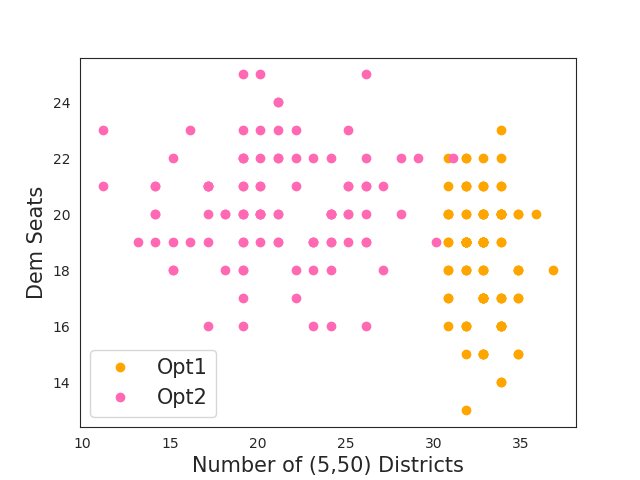}};    
\node at (5,2) {\includegraphics[width=160pt]{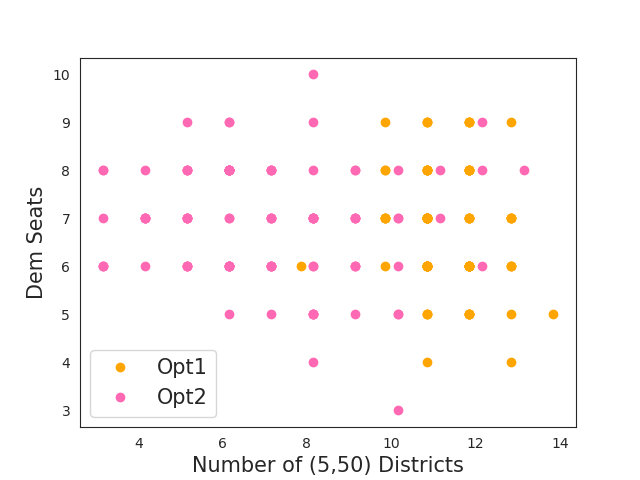}};    
\end{tikzpicture}        
\caption{Number of Democratic seats in the plans found by the two optimization methods plotted against the number of $(5,50)$ districts.  The two types of points are slightly displaced so that they 
are visible in the plot.}
\label{fig:opt}
\end{figure}

Figure \ref{fig:opt} shows the  number of Democratic seats against the number of $(5,50)$ districts found by the two optimization methods. As with our previous examples, 
the effects of promoting plan competitiveness are unpredictable at best, both 
within states or between Congressional and state Senate plans in the same state. 
The plot further highlights that different optimization methods targeting competitiveness can drastically change the partisan character of the plans that are found.  

For Georgia's Congressional plan, higher numbers of competitive districts correlates with Republican seat advantage.  
Wisconsin's Senate plans show the opposite correlation, at least for these optimization techniques,
tending to find more Democratic seats as the number of competitive districts increases. The optimization ensembles also found seats outcomes that did not appear at all in the significantly larger neutral ensembles, which is a reminder that outcomes that are extremely unlikely under partisan-blind circumstances may nonetheless be relatively easy to construct intentionally.  

As with the winnowing results, these optimization runs show that the potentially extreme impacts of enforcing a competitiveness measure depend heavily and unpredictably on  political geography and scale.

\section{Conclusion}
\label{sec:conc}

In the national conversation around redistricting reform, competitiveness is only one among numerous priorities, including traditional districting principles such as population balance or compactness as well as compliance with federal constraints like the Voting Rights Act, that are being debated and sometimes hastily written into law.
Mapping out  the  interactions between various values, criteria, and constraints is becoming increasingly important. 
Data-intensive modeling is indispensable to make informed decisions about these interactions and to identify unintended 
consequences of quantitatively specific rules.  We hope that reform measures for redistricting
will call for this kind of modeling to be performed in every census cycle.

Ensemble analysis---and Markov chain sampling in particular---is a useful tool for exploring these complex spaces and intertwined constraints. The present study sets up a model methodology for surveying the state-specific partisan landscape in advance of 
setting quantitative thresholds and priorities for state reform.
For the specific case of competitiveness, our results show that vote-band rules can be set in a way that does impose 
nontrivial constraints on redistricters, but sometimes with ancillary effects on other partisan metrics.  
There is every reason to expect nontrivial effects on the ability to uphold other priorities---from racial statistics to political boundary integrity---which were not considered here.  
A natural direction for future work is to determine how the operationalizations of state legislative language and other constraints impacts these analyses. Evaluating compliance with the Voting Rights Act of 1965 and the associated Gingles factors, derived from the 1986 Supreme Court ruling in Thornburgh v. Gingles and its successors, provides an example of the complexity of this modeling process and determining interactions between this constraint and competitiveness metrics is an important consideration for future state-specific analyses.

As with any other partisan quantity, securing a high score on a competitiveness metric does not follow effortlessly from the absence of nefarious intent.  
Democratic deliberation is called for both to clarify the type of competitiveness that is sought---more close outcomes? 
fewer landslides?---and to weigh this against other priorities.
As states continue to experiment with their rules and frameworks for redistricting, we hope that statistically robust
ensemble methods become standard tools in the repertoire.

\bibliographystyle{plain}

\bibliography{refs}

\begin{thebibliography}{10}

\bibitem{formula}
Mira Bernstein and Moon Duchin.
\newblock A formula goes to court: Partisan gerrymandering and the efficiency
  gap.
\newblock {\em Notices of the AMS}, 64(9):1020--1024, 2017.

\bibitem{best_authors_2017}
Robin~E. Best, Shawn~J. Donahue, Jonathan Krasno, Daniel~B. Magleby, and
  Michael~D. McDonald.
\newblock Authors' {Response}–-{Values} and {Validations}: {Proper}
  {Criteria} for {Comparing} {Standards} for {Packing} {Gerrymanders}.
\newblock {\em Election Law Journal: Rules, Politics, and Policy},
  17(1):82--84, November 2017.

\bibitem{best_considering_2017}
Robin~E. Best, Shawn~J. Donahue, Jonathan Krasno, Daniel~B. Magleby, and
  Michael~D. McDonald.
\newblock Considering the {Prospects} for {Establishing} a {Packing}
  {Gerrymandering} {Standard}.
\newblock {\em Election Law Journal: Rules, Politics, and Policy}, 17(1):1--20,
  September 2017.

\bibitem{brunell}
Thomas~L. Brunell.
\newblock Rethinking {Redistricting}: {How} {Drawing} {Uncompetitive}
  {Districts} {Eliminates} {Gerrymanders}, {Enhances} {Representation}, and
  {Improves} {Attitudes} toward {Congress}.
\newblock {\em PS: Political Science \& Politics}, 39(1):77--85, January 2006.

\bibitem{WLTSB}
Wisconsin Legislative Technology~Services Bureau.
\newblock Open data portal, 2019.

\bibitem{538}
Aaron Bycoffe, Ella Koeze, David Wasserman, and Julia Wolfe.
\newblock The atlas of redistricting.
\newblock {\em FiveThirtyEight}, 2018.

\bibitem{carson}
Jamie~L. Carson and Michael~H. Crespin.
\newblock The effect of state redistricting methods on electoral competition in
  {U}nited {S}tates {H}ouse of {R}epresentatives races.
\newblock {\em State Politics \& Policy Quarterly}, 4(4):455--469, 2004.

\bibitem{UAGRC}
Utah Automated Geographic~Reference Center.
\newblock Voting precincts, 2019.

\bibitem{cr}
Jowei Chen and Jonathan Rodden.
\newblock Unintentional gerrymandering: Political geography and electoral bias
  in legislatures.
\newblock {\em Quarterly Journal of Political Science}, 8:239--269, 2013.

\bibitem{chikina_practical_2019}
Maria Chikina, Alan Frieze, Jonathan Mattingly, and Wesley Pegden.
\newblock Practical tests for significance in {Markov} {Chains}.
\newblock {\em arXiv:1904.04052 [math, stat]}, April 2019.
\newblock arXiv: 1904.04052.

\bibitem{chikina_assessing_2017}
Maria Chikina, Alan Frieze, and Wesley Pegden.
\newblock Assessing significance in a {Markov} chain without mixing.
\newblock {\em Proceedings of the National Academy of Sciences},
  114(11):2860--2864, March 2017.

\bibitem{UT}
Utah Code.
\newblock Chapter 7 {T}itle 20{A} {C}hapter 19 {P}art 1 {S}ection 103.

\bibitem{cottrell}
David Cottrell.
\newblock Using {Computer} {Simulations} to {Measure} the {Effect} of
  {Gerrymandering} on {Electoral} {Competition} in the {U}.{S}. {Congress}.
\newblock {\em Legislative Studies Quarterly}, 44(3):487--514, 2019.

\bibitem{cottrill}
James~B. Cottrill.
\newblock The {Effects} of {Non}-{Legislative} {Approaches} to {Redistricting}
  on {Competition} in {Congressional} {Elections}.
\newblock {\em Polity}, 44(1):32--50, January 2012.

\bibitem{VA-criteria}
Daryl DeFord and Moon Duchin.
\newblock Redistricting reform in {Virginia}: {Districting} criteria in
  context.
\newblock {\em Virginia Policy Review}, 12(2):120--146, 2019.

\bibitem{PartSymm}
Daryl DeFord, Moon Duchin, Natasha Dhamankar, Varun Gupta, Mackenzie McPike,
  Gabe Schoenbach, and Emily Sim.
\newblock Implementing partisan symmetry: Problems and paradoxes.
\newblock {\em Preprint}, 2019.

\bibitem{hdsr}
Daryl DeFord, Moon Duchin, and Justin Solomon.
\newblock Recombination: A family of {M}arkov chains for redistricting.
\newblock {\em ArXiv:1911.05725}, 2019.

\bibitem{MGGGMA}
Moon Duchin, Taissa Gladkova, Eugene Henninger-Voss, Ben Klingensmith, Heather
  Newman, and Hannah Wheelen.
\newblock Locating the representational baseline: Republicans in
  {M}assachusetts.
\newblock {\em Election Law Journal}, 18, 2019.

\bibitem{redist}
Benjamin Fifield, Michael Higgins, Kosuke Imai, and Alexander Tarr.
\newblock A new automated redistricting simulator using {M}arkov chain {M}onte
  {C}arlo.
\newblock {\em Journal of Computational and Graphical Statistics
  (Forthcoming)}, 2020.

\bibitem{forgette}
Richard Forgette, Andrew Garner, and John Winkle.
\newblock Do {Redistricting} {Principles} and {Practices} {Affect} {U}. {S}.
  {State} {Legislative} {Electoral} {Competition}?
\newblock {\em State Politics \& Policy Quarterly}, 9(2):151--175, 2009.

\bibitem{grainger}
Corbett~A. Grainger.
\newblock Redistricting and {Polarization}: {Who} {Draws} the {Lines} in
  {California}?
\newblock {\em The Journal of Law \& Economics}, 53(3):545--567, 2010.

\bibitem{henderson}
John~A Henderson, Brian Hamel, and Aaron Goldzimer.
\newblock Gerrymandering incumbency: {D}oes non-partisan redistricting increase
  electoral competition?
\newblock {\em Journal of Politics}, 80, 2018.

\bibitem{herschlag_quantifying_2018}
Gregory Herschlag, Han~Sung Kang, Justin Luo, Christy~Vaughn Graves, Sachet
  Bangia, Robert Ravier, and Jonathan~C. Mattingly.
\newblock Quantifying {Gerrymandering} in {North} {Carolina}.
\newblock {\em arXiv:1801.03783 [physics, stat]}, January 2018.
\newblock arXiv: 1801.03783.

\bibitem{herschlag_evaluating_2017}
Gregory Herschlag, Robert Ravier, and Jonathan~C. Mattingly.
\newblock Evaluating {Partisan} {Gerrymandering} in {Wisconsin}.
\newblock {\em arXiv:1709.01596 [physics, stat]}, September 2017.
\newblock arXiv: 1709.01596.

\bibitem{NJ}
New {J}ersey {L}egislature.
\newblock Senate {C}oncurrent {R}esolution 43.

\bibitem{mitedsl}
MIT Election Science~Data Labs.
\newblock Election data, 2019.

\bibitem{GAdata}
Georgia Legislative and Congressional~Reapportionment Office.
\newblock Statewide voting precincts, 2019.

\bibitem{mm1}
Michael~D. McDonald and Robin~E. Best.
\newblock Unfair {Partisan} {Gerrymanders} in {Politics} and {Law}: {A}
  {Diagnostic} {Applied} to {Six} {Cases}.
\newblock {\em Election Law Journal: Rules, Politics, and Policy},
  14(4):312--330, November 2015.

\bibitem{mcghee_rejoinder_2017}
Eric McGhee.
\newblock Rejoinder to “{Considering} the {Prospects} for {Establishing} a
  {Packing} {Gerrymandering} {Standard}”.
\newblock {\em Election Law Journal: Rules, Politics, and Policy},
  17(1):73--82, October 2017.

\bibitem{mggg-states}
MGGG.
\newblock Mggg-{S}tates {G}it{H}ub {R}epository, 2019.

\bibitem{miller}
Peter Miller and Bernard Grofman.
\newblock Redistricting {Commissions} in the {Western} {United} {States}.
\newblock {\em UC Irvine Law Review}, 3(3):637, August 2013.

\bibitem{msced}
Massachusetts~Secretary of~the Commonwealth Elections~Division.
\newblock Election results, 2019.

\bibitem{openprecincts}
Princeton~Gerrymandering Project.
\newblock Openprecincts.org, 2019.

\bibitem{AZ}
Arizona {S}tate {C}onstitution.
\newblock Article {I}{V} {P}art 2 {S}ection 1.

\bibitem{CO}
Colorado {S}tate {C}onstitution.
\newblock Article {V} section 44.

\bibitem{MO}
Missouri {S}tate {C}onstitution.
\newblock Article {I}{I}{I} {S}ection 3.

\bibitem{WA}
Washington {S}tate {C}onstitution.
\newblock Article {I}{I}{I} {S}ection 43.

\bibitem{eg}
Nicholas Stephanopoulos and Eric McGhee.
\newblock Partisan gerrymandering and the efficiency gap.
\newblock {\em University of {C}hicago {L}aw {R}eview}, page 831–900, 2014.

\bibitem{ev}
Ellen Veomett.
\newblock Efficiency gap, voter turnout, and the efficiency principle.
\newblock {\em Election Law Journal}, 17.

\bibitem{gerrychain}
VRDI.
\newblock Gerrychain.
\newblock {\em GitHub Repository: mggg/gerrychain}, 2018.

\bibitem{mm2}
Samuel S.-H. Wang.
\newblock Three {Practical} {Tests} for {Gerrymandering}: {Application} to
  {Maryland} and {Wisconsin}.
\newblock {\em Election Law Journal: Rules, Politics, and Policy},
  15(4):367--384, October 2016.

\bibitem{cvpi}
David Wasserman and Ally Flinn.
\newblock Introducing the 2017 {C}ook {P}olitical {R}eport {P}artisan {V}oter
  {I}ndex.
\newblock 2013.

\bibitem{NY}
New {Y}ork~{S}tate {C}onstitution.
\newblock Article {I}{I}{I} {S}ections 4 and 5.

\end{thebibliography}

\begin{table}[!h]
    \centering
    {\footnotesize
    \begin{tabular}{|c||c|c|c|c|c||c|c|c|c|c|c|}
    \hline
        & \multicolumn{5}{c||}{Democratic seats} & \multicolumn{5}{|c|}{Mean-median} \\
    \hline
 State, Type &  Full& $x_1$ & $x_2$& $x_3$ &$x_4$ &  Full& $x_1$ & $x_2$& $x_3$ &$x_4$  \\
 (Total \#)& Ensemble& &  &  &   &Ensemble&  &  &  &  \\
 \hline
 \hline
   MA Cong (9) & 9.00& 9.00& 9.00& 9.00& ---& 0.021 & 0.021 & 0.015&  0.0094& ---\\
   \hline
     VA Cong (11) &6.35& 6.37& 6.45& 6.66& 6.67&$-0.001$ & 0.001 & 0.002 & 0.003& 0.005\\
   \hline
   WI Cong (8) & 2.66& 2.65& 2.67& 2.80& 3.28& 0.044& 0.044 &0.039 & 0.026& 0.014 \\
   \hline   
    GA Cong (14)  & 4.30& 4.30& 4.30& 4.31& 4.30 & 0.031& 0.028 & 0.024& 0.016& 0.009 \\
   \hline  
   UT Cong (4) & 0.85& 0.85& 0.70& 0.70&---& 0.018 & 0.018& $-0.004$& $-0.014$   &      ---\\
   \hline
   \hline
   MA Sen (40) & 38.36& 38.38& 38.41& 38.48& 38.59  & 0.022 & 0.022& 0.021 &0.020 &0.018 \\
   \hline
   VA Sen (40)& 21.39& 21.38& 21.33& 21.24& 21.16 &0.009 & 0.010& 0.011 & 0.014& 0.016 \\
   \hline
   WI Sen (33) &11.65& 11.65& 11.63& 11.59& 11.46  &0.044& 0.044&0.042& 0.037  &0.034 \\
   \hline
   GA Sen (56)&21.86& 21.86& 21.86& 21.85& 21.82 &0.042 &0.041 &0.038 &0.032&0.024 \\
   \hline
   UT Sen (29) &7.21& 7.22& 7.23& 7.23& 7.22 &0.049& 0.049& 0.049& 0.050& 0.051\\
   \hline
    \end{tabular}
    }
    \caption{Mean partisan statistics (Dem seats, mean-median) in the successively winnowed ensembles shown in Figure~ \ref{fig:onlydemseats}-\ref{fig:mm50state}  for $z=50$.  The $x_i$ values correspond to those in the figures.   For example, the table shows the effects of requiring 6, 7, 8, or 9 competitive districts out of 40 in the MA State Senate, which corresponds to the third figure of the top row of Figure~\ref{fig:onlydemseats}.}
    \label{tab:z50Seats}
\end{table}

\begin{table}[!h]
    \centering
    {\footnotesize
    \begin{tabular}{|c||c|c|c|c|c||c|c|c|c|c|c|}
    \hline
        & \multicolumn{5}{c||}{Democratic seats} & \multicolumn{5}{|c|}{Mean-median} \\
    \hline
 State, Type &  Full& $x_1$ & $x_2$& $x_3$ &$x_4$ &  Full& $x_1$ & $x_2$& $x_3$ &$x_4$  \\
 (Total \#)& Ensemble& &  &  &   &Ensemble&  &  &  &  \\
 \hline
 \hline
   MA Cong (9) & 9.00& 9.00& 9.00& 9.00& 9.00 & 0.021 & 0.021& 0.020& 0.020 & 0.015\\
   \hline
     VA Cong (11) &6.35& 6.45& 6.61& 6.76& 6.97 &$-0.001$& $-0.002$ &$-0.004$   &$-0.004$ &$-0.003$\\
   \hline
   WI Cong (8) & 2.66& 2.65& 2.65& 2.77& 3.24 & 0.044& 0.044 &0.040 &0.027& 0.015\\
   \hline   
    GA Cong (14)  & 4.30& 4.18& 4.04& 3.85& 3.61 & 0.031 & 0.028& 0.026 &0.023  &0.021\\
   \hline  
   UT Cong (4) & 0.85& 0.85& 0.83& 0.86& 0.19 & 0.018 &0.018& 0.021& 0.021& 0.016\\
   \hline
   \hline
   MA Sen (40) & 38.36& 38.36& 38.33& 38.29& 38.25 &0.022& 0.022& 0.021& 0.020& 0.019\\
   \hline
   VA Sen (40)&21.39& 21.44& 21.59& 21.83& 22.12 &0.009 & 0.009 &0.009& 0.008& 0.008\\
   \hline
   WI Sen (33) &11.65& 11.64& 11.63& 11.57& 11.46 &0.044 &0.044&0.042& 0.039&0.035\\
   \hline
   GA Sen (56)&21.86& 21.78& 21.59& 21.28& 20.86 &0.042& 0.041& 0.037& 0.033& 0.029 \\
   \hline
   UT Sen (29) &7.21& 7.20& 7.18& 7.16& 7.11&0.049 &0.047& 0.045 &0.041 &0.036\\
   \hline
    \end{tabular}
    }
    \caption{Mean partisan statistics (Dem seats, mean-median) in the successively winnowed ensembles shown in Figure \ref{fig:onlydemseats}-\ref{fig:mm50state}  for $z=D_0$.  The $x_i$ values correspond to the successively smaller color-coded plots in the figures.   For example, the table shows the effects of requiring 10, 12, 14, or 16 state-typical districts out of 40 in the MA State Senate, which corresponds to the fourth figure of the top row of Figure~\ref{fig:onlydemseats}.}
    \label{tab:zstateSeats}
\end{table}

\end{document}